\documentclass[11pt]{article}
\usepackage{graphicx}
\usepackage{amssymb}
\usepackage{amsmath}
\usepackage{float}
\usepackage{titlesec}
\titleformat{\section}[block]{\bfseries\filcenter}{\thesection.}{0.4em}{}
\titleformat{\subsection}[hang]{\bfseries}{\thesubsection}{1em}{}
\renewcommand{\thesection}{\Roman{section}} 
\renewcommand{\thesubsection}{\Roman{subsection}}
\def\fnote#1#2{\begingroup\def\thefootnote{#1}\footnote{#2}\addtocounter
{footnote}{-1}\endgroup}

\begin{document}

\hfill{UTTG-27-16}

\vspace{20pt}

\begin{center}
{\large {\bf {Gravitational Waves in Cold Dark Matter}}}

\vspace{20pt}

Raphael Flauger\fnote{*}{Electronic address:
flauger@physics.ucsd.edu}\\
{\em Department of Physics, University of California, San Diego \\ La Jolla, CA, 92093}

\vspace{20pt}

Steven Weinberg\fnote{**}{Electronic address:
weinberg@physics.utexas.edu}\\
{\em Theory Group, Department of Physics, University of
Texas\\
Austin, TX, 78712}

\vspace{30pt}

\noindent
{\bf Abstract}
\end{center}

\noindent
We study the effects of cold dark matter on the propagation of gravitational waves of astrophysical and primordial origin. We show that the dominant effect of cold dark matter on gravitational waves from astrophysical sources is a small frequency dependent modification of the propagation speed of gravitational waves. However, the magnitude of the effect is too small to be detected in the near future. We furthermore show that the spectrum of primordial gravitational waves in principle contains detailed information about the properties of dark matter. However, depending on the wavelength, the effects are either suppressed because the dark matter is highly non-relativistic or because it contributes a small fraction of the energy density of the universe. As a consequence, the effects of cold dark matter on primordial gravitational waves in practice also appear too small to be detectable. 

\vfill

\pagebreak

\section{Introduction}

The  direct observation~\cite{Abbott:2016blz} of gravitational waves from distant sources immediately heightened interest in the propagation of these waves from source to detector.     Calabrese, Battaglia, and  Spergel~\cite{Calabrese:2016bnu} considered the future use of gravitational wave source counts as a probe of gravitational wave propagation.  They did not assume any specific model for intervening matter,  supposing instead that by some mechanism the wave intensity falls off as a power of distance.    In contrast, Goswami, Mohanty, and  Prasanna~\cite{Goswami:2016tsu} considered the intervening matter to be an imperfect fluid, using an old result of Hawking~\cite{Hawking:1966qi}, that the intensity of a gravitational wave falls off in an imperfect fluid at a rate $16\pi G\eta$, where $\eta$ is the viscosity.  They set an upper limit on $\eta$ by adopting the estimate of Ref.~\cite{Abbott:2016blz}, that the source is at a distance of 410 Mpc.  This limit would be valid if the source distance really were 410 Mpc, but the source distance was estimated in~\cite{Abbott:2016blz} from the observed signal strength, under the assumption that the gravitational wave is {\em not} damped.   The observations in~\cite{Abbott:2016blz} do not rule out a viscosity greater than the upper bound given in Ref.~\cite{Goswami:2016tsu}; if the viscosity were greater, it would just mean that the distance to the source is less than 410 Mpc.  
In order to use the observed intensity of detected gravitational waves to set an upper limit on the viscosity, we would need an independent measure of the distance of the source, other than the intensity of the gravitational wave.  

But even so, a fundamental question  would remain: Is it reasonable to calculate the effect of cosmic matter on the propagation of gravitational waves by treating this matter as an imperfect fluid?  
It is clear that the treatment of a gas as a fluid, perfect or imperfect, must break down at some sufficiently small collision frequency.  The coefficients of viscosity and heat conduction in the theory of imperfect fluids are proportional to the mean free path, and so would become infinite for zero collision frequency, which is absurd.  The issue whether a particular medium  can be treated as an imperfect fluid, characterized by coefficients of viscosity and heat conduction, depends on the scales of distance and time of the process under study.  As argued briefly in Section III, in the propagation of a gravitational wave through some medium, collisions are effective  only if the mean free path in the medium is smaller than the wavelength.  This is certainly not the case for observed gravitational waves.
The observed wavelengths are in the range of 300 to 15000 km, and there is nothing in interstellar space with free paths that short.  (For hydrogen atoms in our galaxy, with cross sections of the order of a square Angstrom and a density of the order of 1 cm$^{-3}$, the mean free path is of order $10^{11}$ km.  The mean free path of warm ionized gas is somewhat shorter, about $5\times 10^{7}$ km, but still much longer than the observed wavelengths.  Mean free paths are of course longer outside galaxies, and longer for WIMPs everywhere.)  The wavelength of observed gravitational waves is so much smaller  than interstellar and intergalactic mean free paths that it is more appropriate to treat cosmic matter as collisionless than as a fluid, perfect or imperfect.  For this reason, and also with an eye to possible cosmological applications, this paper will  explore the effect  on a gravitational wave of its passage through cold dark matter. 

 The general formalism for calculating the effect of collisionless neutrinos on gravitational waves has already been laid out in \cite{Weinberg:2003ur}.  The perturbation of the neutrinos due to the gravitational wave  was calculated  using the collisionless Boltzmann equation; the result of this calculation was then used to evaluate the effect of the perturbation back on the wave. This formalism  was applied in \cite{Weinberg:2003ur} to cosmological gravitational waves in the radiation-dominated era, in which case the effects were found to be substantial.    
Here we are instead concerned with the effects of massive particles.  Our calculations will follow the same track as in Ref. \cite{Weinberg:2003ur}, but the presence of non-zero mass will make them somewhat more complicated.

In Sections II through V we develop the general formalism for calculating those aspects of the effects of massive collisionless particles on gravitational radiation that are relevant to both astrophysical and cosmological sources. After stating our assumptions in Section II, a general result for the anisotropic inertia in the presence of  massive collisionless matter is given in 
Section III for a general Robertson-Walker scale factor $a(t)$.  In Section IV we apply these results to the case of non-relativistic matter, and give the gravitational wave equation in this case.  Section V deals with  a special cases of relevance to both astrophysical and cosmological sources, of a  wave frequency  much larger than the rate of cosmic expansion.  

  We then consider specific applications.  In Section VI we evaluate the effect of intervening dark matter on the gravitational waves whose detection was reported in \cite{Abbott:2016blz}.  It will be a surprise to no one that the effect turns out to be much too small to be observed. In Section VII we turn to the calculation of the effects of cold dark matter on primordial gravitational waves. Because primordial gravitational waves with wavelengths accessible at interferometers enter the horizon before kinetic decoupling of the dark matter or even when the dark matter is still relativistic, in this section we extend our discussion to include the effects of collisions. We show that the spectrum of primordial gravitational waves in principle contains valuable information about the dark matter like the temperature of kinetic decoupling and the nature of the interactions of dark matter particles. Unfortunately, the effects appear too small to be detectable in the foreseeable future. We summarize our findings in Section VIII.

{\it Added note:} After our paper was nearly finished, we encountered a recent paper \cite{Baym:2017xvh} that covers much the same ground as ours regarding gravitational waves from astrophysical sources, finding as we have that damping of these waves is negligible.  In addition to damping, our discussion pays close attention to the modification of the propagation speed of these waves in cold dark matter, and includes a detailed treatment of the effects of cold dark matter on primordial gravitational waves, which is not considered in \cite{Baym:2017xvh}.  

\section{Assumptions}
  We consider gravitational waves in transverse-traceless gauge in a spatially flat Robertson--Walker background, so that the spacetime line element  takes the form\fnote{*}{We take $i$, $j$, $k$, etc. to run over the spatial coordinate indices 1, 2, 3; repeated indices are summed; and we set the speed of light equal to unity.}
\begin{equation}\label{eq:dt2}
d\tau^2=dt^2-g_{ij}({\bf x},t)\,dx^idx^j\;,
\end{equation}
with 
\begin{equation}\label{eq:gij}
g_{ij}({\bf x},t)=a^2(t)\Big[\delta_{ij}+h_{ij}({\bf x},t)\Big]\,,
\end{equation}
where $|h_{ij}|\ll 1$ and 
\begin{equation}
h_{ii}=0\;,~~~\frac{\partial h_{ij}}{\partial x^i}=0\,.
\end{equation}
Since the background Robertson-Walker metric is invariant under time-independent coordinate-space translations, we can restrict our attention to superpositions of  plane waves with space-dependence 
\begin{equation}
h_{ij}({\bf x},t)\propto e^{i\mathbf{q}\cdot\mathbf{x}}\,,
\end{equation}
where ${\bf q}$ is a time-independent co-moving wave number.  

As is well known, the propagation of the wave represented by $h_{ij}$ is governed by the wave equation
\begin{equation}\label{eq:hijeom}
\ddot{h}_{ij}+\left(\frac{3\dot{a}}{a}\right)\dot{h}_{ij}+\frac{q^2}{a^2}h_{ij}=16\pi G\pi_{ij}\,,
\end{equation}
where $q^2\equiv q_iq_i$, and $\pi_{ij}$ is the anisotropic part of the spatial components of the energy-momentum tensor $T^\mu{}_\nu$:
\begin{equation}
T^i{}_j({\bf x},t)=\pi_{ij}({\bf x},t)+\delta_{ij}\;{\rm terms}\;,~~~~\pi_{ii}({\bf x},t)=0\,.
\end{equation}
We assume that the wave passes through a medium consisting of collisionless particles of mass $m\neq 0$, with an isotropic unperturbed coordinate-space density $4\pi p^2 dp\;\overline{n}(p)$ of particles with $\sqrt{p_ip_i}$ between $p$ and $p+dp$. In particular, our treatment will not include the more familiar effect of gravitational lensing of the gravitational waves by intrinsic density perturbations in the dark matter distribution. Our first task is then to calculate $\pi_{ij}({\bf x},t)$.  The general result for collisionless dark matter found in the following section is given below in Eq.~(\ref{eq:piijcl}). Collisions are included in Section VII.

\section{Calculation of $\pi_{ij}$.}
For a line element of the general form (\ref{eq:dt2}) the four-momentum of a particle of rest-mass $m$ is 
\begin{equation}
p^\mu=m\frac{dx^\mu}{d\tau}\;,
\end{equation}
so 
\begin{equation}
\frac{dx^i}{dt}=p^i/p^0\;,\label{eq:dxdt}
\end{equation}
and 
\begin{equation}
p^0=\sqrt{m^2+g_{ij}p^ip^j}\;.\label{eq:p0}
\end{equation}
It turns out that the covariant components $p_i$ satisfy a simpler equation of motion than the contravariant components
$$
\frac{dp_i}{dt}=\frac{d}{dt}\left(g_{ij}p^j\right)=\frac{\partial g_{ij}}{\partial t}p^j+\frac{\partial g_{ij}}{\partial x^k}\frac{p^kp^j}{p^0}-g_{ij}\Gamma^j_{\mu\nu}\frac{p^\mu p^\nu}{p^0}\;,$$
and therefore for any metric of form (\ref{eq:dt2})
\begin{equation}
\frac{dp_i}{dt}=\frac{1}{2}\frac{\partial g_{kl}}{\partial x^i}\frac{p^k p^\ell}{p^0}\;.
\end{equation}
With the spatial components of the metric of the form (\ref{eq:gij}), this is
\begin{equation}\label{eq:dpdt}
\frac{dp_i}{dt}=\frac{a^2}{2}\frac{\partial h_{kl}}{\partial x^i}\frac{p^k p^\ell}{p^0}=\frac{ia^2}{2} h_{kl}\frac{q_ip^k p^\ell}{p^0}\,,
\end{equation}
so the changes in the covariant components are of first order in the perturbation $h_{ij}$.

For a gas of such particles with $n({\bf p},{\bf x},t)\prod_i dp_i \prod_i dx^i$ particles in a momentum-space volume $\prod_i dp_i$ around ${\bf p}$ and in a coordinate-space volume $\prod_i dx^i$ around ${\bf x}$, the space-components of the energy-momentum tensor are
\begin{equation}\label{eq:Tijn}
T^i{}_j({\bf x},t)=\frac{1}{\sqrt{{\rm 
Det}g({\bf x},t)}}\int d^3p\,n({\bf p},{\bf x},t)\,\frac{p^i({\bf p}, {\bf x},t)p_j}{p^0({\bf p}, {\bf x},t)}\;,
\end{equation}
where $d^3p\equiv \prod_i dp_i $.
The phase space density $n$ is subject to the collisionless Boltzmann equation, which according to Eqs.~(\ref{eq:dxdt}) and (\ref{eq:dpdt}) takes the form
\begin{equation}\label{eq:Beq}
0=\frac{\partial n}{\partial t}+\frac{p^i}{p^0}\frac{\partial n}{\partial x^i}+\frac{ia^2}{2} h_{kl}\frac{q_ip^k p^\ell}{p^0}\frac{\partial n}{\partial p_i}\,.
\end{equation}
We assume that in the absence of the gravitational wave represented by $h_{ij}$ the density $n$  is some function  $\overline{n}\left(\sqrt{p_ip_i}\right)$, which is a trivial solution of Eq,~(\ref{eq:Beq}) for $h_{ij}=0$.  As an initial condition, we suppose that at some initial time $t_1$ the density in the presence of $h_{ij}$  is the same in locally 
Cartesian spatial coordinate  frames: 
\begin{equation}\label{eq:npst1}
n({\bf p},{\bf x},t_1)=\overline{n}\Big(a(t_1)\sqrt{g^{ij}({\bf x},t_1)p_ip_j}\Big)\,.
\end{equation}
To first order in $h_{ij}$, this is
\begin{equation}
n({\bf p},{\bf x},t_1)=\overline{n}(p)-\frac{1}{2}\overline{n}'(p)h_{ij}({\bf x},t_1)p_ip_j/p\,,
\end{equation}
where again $p\equiv \sqrt{p_ip_i}$.  At any later time $t$ there is a dynamical correction $\delta n$ induced by the gravitational wave, so that 
\begin{equation}\label{eq:npxt}
n({\bf p},{\bf x},t)=\overline{n}(p)-\frac{1}{2}\overline{n}'(p)h_{ij}({\bf x},t)p_ip_j/p+\delta n({\bf p},{\bf x},t)\,,
\end{equation}
with initial value $\delta n({\bf p}, {\bf x}, t_1)=0$.  
Since $\partial n/\partial x^i$ is already of first order in $h_{ij}$, in Eq.~(\ref{eq:Beq}) we can use the zeroth order expressions for $p^i$ and $p^0$:
$$p^i=a^{-2}p_i\;,~~~~p^0=\sqrt{m^2+p^2/a^2}\;.$$
Like all other first-order perturbations, $\delta n$ has a space-dependence $\delta n \propto \exp(iq_ix^i)$.  The first-order terms in Eq.~(\ref{eq:Beq}) then give
\begin{eqnarray}\label{eq:Vlasov}
&&\frac{\partial\,\delta n({\bf p}, {\bf x},t)}{\partial t}
+\frac{iq_ip_i}{a^2(t)\sqrt{m^2+p^2/a^2(t)}}\delta n({\bf p}, {\bf x},t)=\frac{ p_kp_l \overline{n}'(p)}{2p }\dot{h}_{kl}({\bf x},t)\;.
\end{eqnarray}

We return to this in detail in Section VII, but let us pause at this point and consider the effect of collisions.  In general, collisions will drive the phase-space distribution back to the equilibrium form (\ref{eq:npst1}), for which  $\delta n=0$, so their effect can be simulated in Eq.~(\ref{eq:Vlasov}) by adding a term $-\Gamma \delta n$ to the right-hand side, where $\Gamma$ is the decay rate of departures from equilibrium in the absence of field perturbations.  Collisions can be ignored if this term is  much less than the transport term in the left-hand side of Eq.~(\ref{eq:Vlasov}) --- that is, if $\Gamma\ll v/\lambda$, where $v=p/a\sqrt{m^2+p^2/a^2}$ is a typical  proper velocity and $\lambda\approx a/q$ is the proper wavelength.  
The decay rate $\Gamma$ varies inversely as the mean free path $\ell$, so on dimensional grounds we expect that $\Gamma\approx v/\ell$.  Hence the condition for neglecting collisions is that $\ell\gg \lambda$.  
As remarked in Section I, this condition is well satisfied for detected gravitational waves.

Returning now to the collisionless Boltzmann equation (\ref{eq:Vlasov}), the solution is
\begin{eqnarray}\label{eq:Vlasovsol}
&&\delta n({\bf p}, {\bf x},t)=\frac{ p_kp_l \overline{n}'(p)}{2p}\int_{t_1}^t dt'\exp\left[-\int_{t'}^t dt''\frac{iq_ip_i}{a^2(t'')\sqrt{m^2+p^2/a^2(t'')}}\right] \dot{h}_{kl}({\bf x},t')\,.
\end{eqnarray}
In calculating the space components (\ref{eq:Tijn}) of the energy-momentum tensor, we use the first-order expressions
\begin{equation}
p^i=a^{-2}[p_i-h_{ik}p_k]\;,~~~~p^0=\sqrt{m^2+p^2/a^2}-\frac{h_{kl}p_kp_l}{2a^2\sqrt{m^2+p^2/a^2}}\;,
\end{equation}
and Eqs.~(\ref{eq:npxt}) and (\ref{eq:Vlasovsol}).  To first order in $h_{ij}$ the spatial components of the energy-momentum tensor are then
\begin{eqnarray}
&&\hskip -2cm T^i{}_j({\bf x},t)=\frac{1}{a^5(t)}\int d^3p\;\overline{n}(p)\,\Bigg[\frac{p_ip_j}{\sqrt{m^2+p^2/a^2(t)}}
\nonumber\\
&&\hskip 2.8cm-\frac{h_{ik}({\bf x},t)p_kp_j}{\sqrt{m^2+p^2/a^2(t)}}+\frac{p_ip_jp_kp_l h_{kl}({\bf x},t)}{2a^2(t)(m^2+p^2/a^2(t))^{3/2}}\Bigg]\nonumber\\
&&-\frac{1}{2a^5(t)}\int d^3p\;\overline{n}'(p)\,\frac{p_ip_jp_kp_lh_{kl}({\bf x},t)}{p\sqrt{m^2+p^2/a^2(t)}}\nonumber\\&&
+\frac{1}{a^5(t)}\int d^3p\; \overline{n}'(p)\frac{p_ip_jp_kp_l}{2p\sqrt{m^2+
p^2/a^2(t)}}\nonumber\\
&&\hskip 2cm\times\int_{t_1}^{t}dt'\dot{h}_{kl}({\bf x},t')\exp\left[-\int_{t'}^tdt''\frac{iq_ip_i}{a^2(t'')\sqrt{m^2+
p^2/a^2(t'')}}\right]\label{eq:Th1ij}\,.
\end{eqnarray}
The next-to-last term of Eq.~(\ref{eq:Th1ij}) can be calculated by setting $\overline{n}'(p)p_i/p = \partial \overline{n}(p)/\partial p_i$ and integrating by parts in momentum space.  In this way we find that all the terms in 
Eq.~(\ref{eq:Th1ij})  cancel, except for a term proportional to $\delta_{ij}$  
and the last term in 
Eq.~(\ref{eq:Th1ij}):
\begin{eqnarray}\label{eq:Th1ij2}
&&\hskip -1cm T^i{}_j({\bf x},t)=\frac{1}{a^5(t)}\int d^3p\; \overline{n}'(p)\frac{p_ip_jp_kp_l}{2p\sqrt{m^2+p^2/a^2(t)}}\nonumber\\&&\hskip 2cm\times\int_{t_1}^{t}dt'\dot{h}_{kl}({\bf x},t')\exp\left[-i\int^t_{t'}dt''\frac{q_ip_i}{a^2(t'')\sqrt{m^2+p^2/a^2(t'')}}\right]\nonumber\\&&\hskip 1cm +\delta_{ij}\;{\rm terms}\;.
\end{eqnarray}

The momentum space volume element in  Eq.~(\ref{eq:Th1ij2}) may be written as $d^3p=p^2\,dp\,dz\,d\varphi$, where $z=q_ip_i/qp$ is the cosine of the angle between the wave vector ${\bf q}$ and the momentum ${\bf p}$, and $\varphi$ is the azimuthal angle of the momentum around the wave vector.  
The integral of $p_ip_jp_kp_l/p^4$ over $\varphi$ must take the form of a linear combination of symmetric terms formed from Kronecker deltas and $\hat{q}\equiv {\bf q}/q$, with coefficients that depend only on $z$
\begin{eqnarray*}
&&\hskip -1.75cm\int_0^{2\pi}d\varphi\; p_ip_jp_kp_l/p^4=A(z)\hat{q}_i\hat{q}_j\hat{q}_k\hat{q}_l\\
&&\hskip 1.7cm+B(z)[\hat{q}_i\hat{q}_j\delta_{kl}+\hat{q}_i\hat{q}_k\delta_{jl}+\hat{q}_i\hat{q}_l\delta_{jk}+\hat{q}_j\hat{q}_k\delta_{il}+\hat{q}_j\hat{q}_l\delta_{ik}+\hat{q}_k\hat{q}_l\delta_{ij}]\\
&&\hskip 1.7cm+ C(z)[\delta_{ij}\delta_{kl}+\delta_{ik}\delta_{jl}+\delta_{il}\delta_{jk}]\,.
\end{eqnarray*}
Because $h_{kl}$ is transverse and traceless, terms proportional to $\hat{q}_k$ or $\hat{q}_l$ or $\delta_{kl}$ do not contribute in Eq.~(21), so all we need is $C(z)$, which by taking various contractions is easily calculated to be $C(z)=\pi(1-z^2)^2/4$.  Discarding terms proportional to $\delta_{ij}$, Eq.~(\ref{eq:Th1ij2}) finally gives the anisotropic stress tensor for collisionless particles
\begin{eqnarray}
&&\hskip -2cm\pi_{ij}({\bf x},t)=\frac{\pi}{4a^5(t)}\int_0^\infty p^5\,dp\; \frac{\overline{n}'(p)}{\sqrt{m^2+p^2/a^2(t)}}\int_{-1}^{+1} (1-z^2)^2\,dz\nonumber\\
&&\hskip 2cm\times\int_{t_1}^{t}dt'\dot{h}_{ij}({\bf x},t')\exp\left[-
\int^t_{t'}dt''\frac{iqpz}{a^2(t'')\sqrt{m^2+
p^2/a^2(t'')}}\right]\;.\label{eq:piijcl}
\end{eqnarray}
This is traceless and transverse because $h_{ij}$ is.

As a check on Eq.~(\ref{eq:piijcl}), let's briefly consider the special case of massless collisionless particles such as neutrinos, or at any rate particles that have $p/a(t)\gg m$ during the period of interest.  Here Eq.~(\ref{eq:piijcl}) becomes
\begin{eqnarray*}
&&\pi_{ij}({\bf x},t)=\frac{\pi}{4a^4(t)}\int_0^\infty p^4\,dp\; \overline{n}'(p)\int_{-1}^{+1} (1-z^2)^2\,dz\int_{t_1}^{t}dt'\dot{h}_{ij}({\bf x},t')\exp\left[-
\int^t_{t'}dt''\frac{iqz}{a(t'')}\right]\;.
\end{eqnarray*}
The argument of the exponential does not depend on $p$, so if we integrate over $p$ by parts we have
\begin{eqnarray*}
&&\pi_{ij}({\bf x},t)=-\frac{\pi}{a^4(t)}\int_0^\infty p^3\,dp\;\overline{n}(p)\int_{-1}^{+1} (1-z^2)^2\,dz\int_{t_1}^{t}dt'\dot{h}_{ij}({\bf x},t')\exp\left[-
\int^t_{t'}dt''\frac{iqz}{a(t'')}\right]\;.
\end{eqnarray*}
To zeroth order in $h_{ij}$, the the proper volume of a coordinate space volume $d^3x$ is $a^3d^3x$, and the energy of a massless particle is given by Eq.~(\ref{eq:p0}) as $p^0=a^{-1}\sqrt{p_ip_i}=a^{-1}p$, so the total energy  per proper volume is 
$$
\rho(t)=\int d^3p\;\overline{n}(p)p/a^4(t)= 4\pi\int_0^\infty  p^3\,dp\;\overline{n}(p)/a^4(t)\;.
$$
For $m=0$ Eq.~(\ref{eq:piijcl}) therefore gives:
\begin{eqnarray*}
&&\pi_{ij}({\bf x},t)=-\frac{\rho(t)}{4}\int_{-1}^{+1} (1-z^2)^2\,dz\int_{t_1}^{t}dt'\dot{h}_{ij}({\bf x},t')\exp\left[-
\int^t_{t'}dt''\frac{iqz}{a(t'')}\right]\;,
\end{eqnarray*}
which is the same result as given for neutrinos by Eqs.~(\ref{eq:npxt}) and (\ref{eq:Vlasov}) of reference \cite{Weinberg:2003ur}.

\vspace{10pt}

\section{Non-relativistic matter}

For a general non-zero particle mass $m$, our result (\ref{eq:piijcl}) for $\pi_{ij}$ is much more complicated than for $m=0$.  We can regain some of the simplicity of the zero mass case by specializing to the opposite limit, of non-relativistic matter.  We will now  assume (as is likely for dark matter) that the  matter through which the gravitational wave passes is non-relativistic, in the sense that $\overline{n}(p)$ is non-negligible only for $p$ small enough so that
\begin{equation}\label{eq:nonrel}
p/a(t')\ll m\,,
\end{equation}
over the whole time $t'$ from emission of the gravitational wave at $t'=t_1$ to direct or indirect detection of the gravitational wave at $t'=t$.  
Then Eq.~(\ref{eq:piijcl}) becomes
\begin{eqnarray}\label{eq:piij}
&&\hskip -2cm\pi_{ij}({\bf x},t)=\frac{\pi}{4a^5(t)m}\int_0^\infty p^5\,dp\; \overline{n}'(p)\int_{-1}^{+1} (1-z^2)^2\,dz\nonumber\\
&&\hskip 1.5cm\times\int_{t_1}^{t}dt'\dot{h}_{ij}({\bf x},t')\exp\left[-i(p/m)z
\int^t_{t'}dt''\frac{q}{a^2(t'')}\right]\;.
\end{eqnarray}

If the dark matter particles move less than the wavelength of the mode between $t''=t_1$ to $t''=t$, the argument of the exponential in Eq.~(\ref{eq:piij}) is small.  The integral over $t'$ is then trivial; the integral of $(1-z^2)^2$ over $z$ just gives a factor 16/15; and the integral over $ p$ can be done by parts, so that
\begin{equation}\label{eq:piij2}
\pi_{ij}({\bf x},t)=-\frac{2{\cal E}}{3a^5(t)}\Big[h_{ij}({\bf x},t)-h_{ij}({\bf x},t_1)\Big]\;,
\end{equation}
where
\begin{equation}\label{eq:E}
{\cal E}\equiv \int_0^\infty 4\pi p^2\,\overline{n}(p)\,dp\times \frac{p^2}{2m}\;.
\end{equation}
(Note that ${\cal E}/a^5(t)$ is the proper kinetic energy density at time $t$.)  The wave equation (\ref{eq:hijeom}) can thus be written as
\begin{equation}\label{eq:eom_hij}
\ddot{h}_{ij}({\bf x},t)+3\left(\frac{\dot{a}(t)}{a(t)}\right)\dot{h}_{ij}({\bf x},t)+\omega^2(t)h_{ij}({\bf x},t)=\frac{32\pi G{\cal E}}{3a^5(t)}h_{ij}({\bf x},t_1)\;,
\end{equation}
where\footnote{Notice that the modification of the dispersion relation comes with definite sign, and that the phase velocity is greater than the speed of light so that there can be no gravitational Cherenkov radiation.} 
\begin{equation}\label{eq:omq}
\omega^2(t)\equiv \frac{q^2}{a^2(t)}+\frac{32\pi G{\cal E}}{3a^5(t)}\,.
\end{equation}

   In general, matters are more complicated.  The  non-relativistic assumption  (\ref{eq:nonrel}) does not automatically allow us to  set the argument of the exponential in Eq.~(\ref{eq:piij}) equal to zero. Even non-relativistic particles will travel a distance large compared to the wavelength if given enough time, making the argument of the exponential in Eq.~(\ref{eq:piij})  much larger than unity.   We will see in Section V that this is likely the case for the gravitational waves reported in \cite{Abbott:2016blz}. However, under the relativistic assumption the rate of oscillation of the exponential in Eq.~(\ref{eq:piij}) is much smaller the rate of oscillation of $h_{ij}$, which is of order $q/a$. So we can take the $t'$-derivative in Eq.~(\ref{eq:piij}) to act on the whole integrand of the integral over $t'$:
\begin{equation}\label{eq:ibp}
\dot{h}_{ij}({\bf x},t')\exp\left[-
\int^t_{t'}dt''\frac{iqpz}{ma^2(t'')}\right]\simeq \frac{\partial}{\partial t'}\left\{h_{ij}({\bf x},t')\exp\left[-\int^t_{t'}dt''\frac{iqpz}{ma^2(t'')}\right]\right\}\;.
\end{equation}
The integral over $t'$ is then trivial, and we find
\begin{eqnarray}\label{eq:piijsimp}
&&\hskip -2cm\pi_{ij}({\bf x},t)\simeq \frac{\pi}{4a^5(t)m}\int_0^\infty p^5\,dp\; \overline{n}'(p)\int_{-1}^{+1} (1-z^2)^2\,dz\nonumber\\&&\hskip 2cm\times\left\{h_{ij}({\bf x},t)-h_{ij}({\bf x},t_1)\exp\left[-i
\int^t_{t_1}dt''\frac{qpz}{ma^2(t'')}\right]\right\}\;.
\end{eqnarray}

To see what sort of error is introduced in this approximation,  consider for a moment a case in which the original $t'$ integral can be done explicitly for general mass without the approximation (\ref{eq:ibp}) .   Suppose that 
$a(t)$ is a constant, which can be taken as $a(t)=1$, and suppose that the gravitational wave has a simple-harmonic time-dependence
$$
h_{ij}({\bf x},t)=C_{ij}\exp\Big(i{\bf q}\cdot{\bf x}\Big)\exp\Big(\pm i\omega(t-t_1)\Big)\;,
$$
with $C_{ij}$ constant, and $\omega$  a constant frequency, of order $q$.    The integral over $t'$ in Eq.~(\ref{eq:piij}) is then straightforward
\begin{eqnarray*}
&&\hskip -1cm\pi_{ij}({\bf x},t)=\frac{\pi}{4}\int_0^\infty p^4\,v\,dp\; \overline{n}'(p)\int_{-1}^{+1} (1-z^2)^2\,dz\nonumber\\&&~~~~~~\times C_{ij}\exp\Big(i{\bf q}\cdot{\bf x}\Big)\frac{\omega}{\omega\mp  vzq}\Bigg[\exp\Big(\pm i\omega(t-t_1)\Big)-\exp\Big( -iqvz(t-t_1)\Big)\Bigg]\nonumber\,,
\end{eqnarray*}
where $v\equiv p/m$.  
Comparison with Eq.~(\ref{eq:piijsimp}) shows that in this case,  the approximation (\ref{eq:ibp}) just amounts to supposing that $v$ is small enough to allow us to replace the factor $\omega/(\omega\mp vzq)$ with  unity.
 
Coming back  to Eq.~(\ref{eq:piijsimp}),  the wave equation (\ref{eq:hijeom}) may now be written
\begin{equation}\label{eq:hijSij}
\ddot{h}_{ij}({\bf x},t)+3\left(\frac{\dot{a}(t)}{a(t)}\right)\dot{h}_{ij}({\bf x},t)+\omega^2(t)h_{ij}({\bf x},t)=S_{ij}({\bf x},t)\;,
\end{equation}
where again
\begin{equation}
\omega^2(t)=\frac{q^2}{a^2(t)}+\frac{32\pi G\cal{E}}{3a^5(t)}\;,~~~{\cal E}\equiv \int_0^\infty 4\pi p^2\,\overline{n}(p)\,dp\times\frac{p^2}{2m}\,,
\end{equation}
and $S_{ij}$ is $16\pi G$ times the second term in $\pi_{ij}$:
\begin{eqnarray}
&&\hskip -1cm S_{ij}({\bf x},t)\equiv -h_{ij}({\bf x},t_1)\frac{4\pi^2 G}{a^5(t)m}\int_0^\infty p^5\,dp\; \overline{n}'(p)\int_{-1}^{+1} \!dz\,(1-z^2)^2\exp\left[-i\int^t_{t_1}dt''\frac{qpz}{ma^2(t'')}\right]\;.
\end{eqnarray}
We write the wave equation in this form because the right-hand side $S_{ij}$ is a transient that goes to zero exponentially with increasing $t$ after the dark matter particles have traveled a distance larger than the wavelength of the mode. More concretely, if for some $t_2$ we have $qp/m\int_{t_1}^{t_2}dt''/a^2(t'')\gg 1$; then for any smooth density function  $\overline{n}(p)$ of $p$, $S_{ij}$ becomes exponentially small for $t>t_2$.

To illustrate this, let us take  $\overline{n}(p)$  to have the Maxwell-Boltzmann form 
\begin{equation}\label{eq:Bdist}
\overline{n}(p)=N\exp(- p^2/2P^2)\,,
\end{equation}
with $N$ and $P$ any positive constants.  The $z$ and $p$ integrals are then straightforward, and we find that the wave equation (\ref{eq:hijSij}) takes the form
\begin{equation}\label{eq:hijBdist}
\ddot{h}_{ij}({\bf x},t)+3\left(\frac{\dot{a}(t)}{a(t)}\right)\dot{h}_{ij}({\bf x},t)+\omega^2(t)h_{ij}({\bf x},t)
= h_{ij}({\bf x},t_1)\frac{32\pi G{\cal E}}{3a^5(t)}
\exp\left[-\frac{\overline{v^2}}{2}\left(\int_{t_1}^t \frac{q\,dt'}{a^2(t')}\right)^2\right]\;,
\end{equation}
where ${\cal E}$ is again given by Eq.~(\ref{eq:E}), and $\overline{v^2}=P^2/m^2$ is the mean square coordinate velocity for the distribution (\ref{eq:Bdist}).  Our assumption that $\overline{v^2}/a^2(t'')\ll 1$ makes the argument of the exponential in Eq.~(\ref{eq:hijBdist}) negligible  in the case of few oscillations, so that in this case the wave equation (\ref{eq:hijBdist}) agrees with our earlier result (\ref{eq:eom_hij}), and we can take Eq.~(\ref{eq:eom_hij}) as a fair approximation to the wave equation for all times.  But $S_{ij}({\bf x},t)$ is exponentially small for late times when the dark matter particles have traveled far compared to the wavelength of the mode and the number of oscillations becomes so large that
$$
\sqrt{\overline{v^2}}\int_{t_1}^t \frac{q\,dt'}{a^2(t')}\gg 1\;.
$$
At these late times, the memory of the gravitational field at the time of emission in the distribution of momenta is erased, and the wave equation (\ref{eq:hijBdist}) simply becomes
\begin{equation}\label{eq:hijhom}
\ddot{h}_{ij}({\bf x},t)+3\left(\frac{\dot{a}(t)}{a(t)}\right)\dot{h}_{ij}({\bf x},t)+\omega^2(t)h_{ij}({\bf x},t)=0\;.
\end{equation}
But to find the coefficients of the two independent solutions of the homogeneous equation (\ref{eq:hijhom}) we need to use the inhomogeneous wave equation,  Eq.~(\ref{eq:hijBdist}).

\vspace{10pt}

\section{Short Wavelengths}

It is not possible to find  analytic solutions of either Eq.~(\ref{eq:hijBdist}) or Eq.~(\ref{eq:hijhom}) for an arbitrary time-dependence of the  Robertson--Walker scale factor $a(t)$.  But we can find solutions when the frequency $\omega(t)$ is much larger than the fractional time-dependence $H(t)=\dot{a}(t)/a(t)$ of the scale factor, and hence also much larger than the fractional time-dependence of $\omega(t)$ itself.  This of course includes the case of constant $a(t)$, which is a good approximation for the gravitational waves reported in \cite{Abbott:2016blz}, and to which we shall return in Section VI.  

In the short-wavelength case, the familiar WKB approximation (neglecting second time derivatives of the coefficients of the cosine or sine) yields  approximate solutions of the homogeneous equation (\ref{eq:hijhom}), with time-dependence   
$$
a^{-3/2}(t)\,\omega^{-1/2}(t)\times\begin{array}{c}\cos\left[\int^t \omega(t')\,dt'\right]\\ \sin\left[\int^t \omega(t')\,dt'\right]\,. \end{array}
$$
Knowing these homogeneous solutions, it is easy to construct a Green's function that allows us to solve the inhomogeneous equation (\ref{eq:hijBdist})  
$$ G(t,t')\equiv \frac{a^{3/2}(t')\omega^{-1/2}(t')}{a^{3/2}(t)\omega^{1/2}(t)}\sin\left[\int_{t'}^t\omega(t'')\,dt''\right]\theta(t-t')\,,$$
for which, within the WKB approximation,
$$\left[\frac{d^2}{dt^2}+3\left(\frac{\dot{a}(t)}{a(t)}\right)\frac{d}{dt}+\omega^2(t)\right]G(t,t')=\delta(t-t')\;.$$
The general solution of Eq.~(\ref{eq:hijBdist}) is therefore
\begin{eqnarray}\label{eq:hijBdistsol}
&&\hskip -1cm h_{ij}({\bf x},t)=h^{(0)}_{ij}({\bf x},t)\nonumber\\
&&\hskip .5cm  +\frac{32\pi G {\cal E}}{3}h_{ij}({\bf x},t_1)\int_{t_\star}^{t}\frac{dt'}{a^5(t')\omega(t')}\frac{a^{3/2}(t')\omega^{1/2}(t')}{a^{3/2}(t)\omega^{1/2}(t)}
\sin\left[\int_{t'}^t\omega(t'')\,dt''\right]
\nonumber\\&&\hskip 7cm\times \exp\left[-\frac{\overline{v^2}}{2}\left(\int_{t_1}^{t'} \frac{q\,dt''}{a^2(t'')}\right)^2\right]\;,
\end{eqnarray}
where $h^{(0)}_{ij}({\bf x},t)$ is some  solution of the homogeneous equation (\ref{eq:hijhom}).  The lower bound $t_\star$ on the integral over $t'$ is arbitrary, because the difference in the integral between two possible choices of $t_\star$ is a solution of the homogeneous equation (\ref{eq:hijhom}), and so far $h^{(0)}$ is an arbitrary solution of the homogeneous equation.  The one condition that must be satisfied by $t_\star$ is that the WKB approximation must be valid from $t_\star$ to $t$.  This may or may not allow us to choose $t_\star=t_1$, depending on the context.  Whatever we choose for $t_\star$, the inhomogeneous term in Eq.~(\ref{eq:hijBdistsol}) and its first time-derivative both vanish for $t=t_\star$, so the homogeneous term by itself must satisfy the initial conditions at $t=t_\star$, and therefore takes the form
\begin{eqnarray}\label{eq:hijhomsol}
&&\hskip -1.5cm h^{(0)}_{ij}({\bf x},t)=\frac{a^{3/2}(t_\star)\omega^{1/2}(t_\star)}{a^{3/2}(t)\omega^{1/2}(t)}\Bigg[h_{ij}({\bf x},t_\star)\cos\left(\int^t_{t_\star} \omega(t')\,dt'\right)\nonumber\\&&\hskip 3cm
+\dot{h}_{ij}({\bf x},t_\star)\omega^{-1}(t_\star)\sin\left(\int^t_{t_\star} \omega(t')\,dt'\right)\Bigg]\;.
\end{eqnarray}

We are now in a position to evaluate the coefficients of the solutions of the homogeneous equation  after many oscillations.  We write the argument of the  sine in Eq.~(\ref{eq:hijBdistsol}) as
$$\int_{t'}^t\omega(t'')\,dt''=\int_{t_\star}^t\omega(t'')\,dt''-\int_{t_\star}^{t'}\omega(t'')\,dt''\;.$$
Then Eqs.~(\ref{eq:hijBdistsol}) and (\ref{eq:hijhomsol}) become
\begin{eqnarray}\label{eq:hijinhomsol}
&&\hskip -1cm h_{ij}({\bf x},t)=\frac{a^{3/2}(t_\star)\omega^{1/2}(t_\star)}{a^{3/2}(t)\omega^{1/2}(t)}\Bigg[\cos\left(\int_{t_\star}^t \omega(t'')dt''\right)\Big(h_{ij}({\bf x},t_\star)+A(t)h_{ij}({\bf x},t_1)\Big)\nonumber\\
&&\hskip 2cm+\sin\left(\int_{t_\star}^t \omega(t'')dt''\right)\left(\omega^{-1}(t_\star)\dot{h}_{ij}({\bf x},t_\star)
+B(t)h_{ij}({\bf x},t_1)\right)\Bigg]\;,\label{eq:WKBV}
\end{eqnarray}
where
\begin{eqnarray}
&&\hskip -0.5cm A(t)=-\frac{32\pi G {\cal E}}{3}\int_{t_\star}^{t}\frac{dt'}{a^5(t')}\frac{a^{3/2}(t')\omega^{-1/2}(t')}{a^{3/2}(t_\star)\omega^{1/2}(t_\star)}
\sin\left[\int_{t_\star}^{t'}\omega(t'')\,dt''\right]\nonumber\\&&\hskip 5cm\times\exp\left[-\frac{\overline{v^2}}{2}\left(\int_{t_1}^{t'} \frac{q\,dt''}{a^2(t'')}\right)^2\right]\,,~~~~~~~ \nonumber \\
&&\hskip -0.5cm B(t)=\frac{32\pi G {\cal E}}{3}\int_{t_\star}^{t}\frac{dt'}{a^5(t')}\frac{a^{3/2}(t')\omega^{-1/2}(t')}{a^{3/2}(t_\star)\omega^{1/2}(t_\star)}
\cos\left[\int_{t_\star}^{t'}\omega(t'')\,dt''\right]\nonumber\\&&\hskip 5cm\times\exp\left[-\frac{\overline{v^2}}{2}\left(\int_{t_1}^{t'} \frac{q\,dt''}{a^2(t'')}\right)^2\right]\label{eq:Bint}\,.
\end{eqnarray}
If at some time  $t'=t_2$ the argument of the exponentials in Eq.~(\ref{eq:Bint}) becomes much larger than unity, the integrals of $t'$ are effectively cut off for $t'>t_2$, and  $A(t)$ and $B(t)$ approach finite $t$-independent values for $t>t_2$.  The solution (\ref{eq:hijinhomsol}) then becomes a linear combination of solutions of the homogeneous equation.
\begin{eqnarray}
&&\hskip -1cm h_{ij}({\bf x},t)\rightarrow \frac{a^{3/2}(t_\star)\omega^{1/2}(t_\star)}{a^{3/2}(t)\omega^{1/2}(t)}\Bigg[\cos\left(\int_{t_\star}^t \omega(t'')dt''\right)\Big(h_{ij}({\bf x},t_\star)+A(\infty)h_{ij}({\bf x},t_1)\Big)\nonumber\\&&
\hskip 3.5cm+\sin\left(\int_{t_\star}^t \omega(t'')dt''\right)\left(\omega^{-1}(t_\star)\dot{h}_{ij}({\bf x},t*)
+B(\infty)h_{ij}({\bf x},t_1)\right)\Bigg]\;.
\end{eqnarray}

\section{Observed Gravitational Waves}

\vspace{10pt}

As a first application of our results for $m\neq 0$, let us consider the effect of intervening dark matter on observed gravitational waves~\cite{Abbott:2016blz}, believed to be produced by coalescing black holes.  Since the source of these waves is at a fairly small redshift $z<0.1$, we can greatly simplify our calculations by taking the Robertson--Walker scale  factor $a(t)$ to be constant during the time elapsed from production to detection of the waves.  Without loss of generality we can normalize our spatial coordinates so that $a(t)=1$.

For $a(t)=1$, the gravitational wave equation (\ref{eq:hijBdist}) in the presence of collisionless non-relativistic matter here takes the form
\begin{equation}\label{eq:hijobs}
\ddot{h}_{ij}({\bf x},t)+\omega^2h_{ij}({\bf x},t)
= h_{ij}({\bf x},t_1)\frac{32\pi G{\cal E}}{3}
\exp\left[-\frac{\overline{v}^2q^2(t-t_1)^2}{2}\right]\;,
\end{equation}
where now the frequency (\ref{eq:omq}) is a constant
\begin{equation}\label{eq:omqobs}
\omega^2=q^2+\Omega^2\;,~~~~\Omega^2=\frac{32\pi G {\cal E}}{3}\,.
\end{equation}
and ${\cal E}$ is the proper density of kinetic energy.  

With $a(t)$ constant we can use the results of the previous section, with no need for the WKB approximation.  Since we are not relying here on the WKB approximation, there is no obstacle to taking the lower bound $t_\star$ in Eqs.~(\ref{eq:hijinhomsol}) and (\ref{eq:Bint}) to be equal to the emission time $t_1$.  The solution  (\ref{eq:hijinhomsol}) of Eq.~(\ref{eq:hijobs}) is now exact, and takes the form
\begin{eqnarray}
&&h_{ij}({\bf x},t)=\cos\left(\omega(t-t_1)\right)\left(1+A(t)\right)h_{ij}({\bf x},t_1)\nonumber\\
&&\hskip 1.5cm +\sin\left( \omega(t-t_1)\right)\left(\omega^{-1}\dot{h}_{ij}({\bf x},t_1)
+B(t)h_{ij}({\bf x},t_1)\right)\;,\label{eq:hsolobs}
\end{eqnarray}
where
\begin{eqnarray}
&&A(t)=-\frac{32\pi G {\cal E}}{3\omega}\int_{t_1}^{t}dt'
\sin\left[\omega (t'-t_1)\right]\exp\left[-\frac{\overline{v^2}q^2(t'-t_1)^2}{2}\right]\,, \label{eq:Aintobs}\\
&&B(t)=\frac{32\pi G {\cal E}}{3\omega}\int_{t_1}^{t}dt'
\cos\left[\omega (t'-t_1)\right]\exp\left[-\frac{\overline{v^2}q^2(t'-t_1)^2}{2}\right]\label{eq:Bintobs}\,.
\end{eqnarray}
The gravitational waves with the lowest  observed frequencies have wavelength about  15000 km, so if their source is at a distance 410 Mpc,\footnote{The values here correspond to those in reference \cite{Abbott:2016blz} because much of the paper was written shortly after the discovery of gravitational waves. The conclusions remain the same for the more recent observations of gravitational wave events.} the quantity $q(t-t_1)$ is of order $5\times 10^{18}$.  Hence the argument  of the exponentials in Eqs.~(\ref{eq:Aintobs}) and (\ref{eq:Bintobs})  is already much larger than unity even for $t'$ much less than $t$, provided that the rms velocity of the dark matter is much larger than $2\times 10^{-19}c$, which we shall assume to be the case.  In this case the dark matter particles travel a distance long compared to the wavelength of the gravitational wave, and the exponentials in Eqs.~(45) and (46) therefore cut off the integrals already for $t'$ much less than $t$, and we can take $t=\infty$ in $A(t)$ and $B(t)$.  The integral for $B(\infty)$ is easy
\begin{equation}
B(\infty)=\frac{32\pi G {\cal E}}{3\omega q}\sqrt{\frac{\pi}{2\overline{v^2}}}\exp\left[-\frac{\omega^2}{2\overline{v^2}q^2}\right]\;.
\end{equation}
The integral for $A(\infty)$ is more complicated.  It can be expressed in terms of a confluent hypergeometric function of the first kind
\begin{equation}
A(\infty)=-\frac{32\pi G {\cal E}}{3 q^2\overline{v^2}}\exp\left(-\frac{\omega^2}{2\overline{v^2}q^2}\right)\;{}_1F_1\left(\frac{1}{2},\frac{3}{2},\frac{\omega^2}{2\overline{v^2}q^2}\right)\;,
\end{equation}
with \cite{Grad}
\begin{equation}{}_1F_1\left(\frac{1}{2},\frac{3}{2},z\right)=2^{-3/2}\int_{-1}^1(1+t)^{-1/2}\exp\left(z(1+t)/2\right)dt\,.
\end{equation}
Of particular interest is the limit $\overline{v^2}\rightarrow 0$, with $\omega/q$ of order unity.  In this limit $B(\infty)$ is exponentially small, while $A(\infty)\rightarrow -\Omega^2/\omega^2$, a result that can be obtained more simply by writing $\sin\omega (t-t_1)$ in Eq.~(\ref{eq:Aintobs}) as $(1/\omega)(d/dt)\cos\omega (t-t_1)$ and integrating by parts.  In this limit Eq.~(\ref{eq:hsolobs}) becomes
\begin{eqnarray}\label{eq:hijobslimit}
&&h_{ij}({\bf x},t)=\cos\left(\omega(t-t_1)\right)\left(1-\frac{\Omega^2}{\omega^2}\right)h_{ij}({\bf x},t_1)+\omega^{-1}\sin\left( \omega(t-t_1)\right)\dot{h}_{ij}({\bf x},t_1)\;.
\end{eqnarray}

One effect of the modified relation (\ref{eq:omqobs}) between $q$ and $\omega$ is a frequency-dependence of the group velocity
$$
v_g=\frac{\partial \omega}{\partial q}=\sqrt{1-\Omega^2/\omega^2}\,.
$$
After the gravitational wave has traveled for a distance $D$,  two components of the wave of different frequency  will arrive at times separated by
$\Delta t=D\Delta(1/v_g)$.  In addition to the shift in frequency, the presence of the correction term proportional to $\Omega^2/\omega^2$ in the relation (\ref{eq:hijobslimit}) between the observed gravitational wave and the initial conditions leads to some distortion of the gravitational waveform.

But if dark matter is composed of WIMPs, these effects are extremely small.  Even if we were to suppose that dark matter particles have moderate velocities, and dominate the cosmic energy density $\rho_0$, the quantity $\Omega$ would be no greater than $H_0=\sqrt{8\pi G\rho_0/3}$, which of course is tiny compared with $\omega$ for observed gravitational waves, so $\Omega^2/\omega^2$ is negligible.  The correction to the group velocity has a larger effect, but one that is still very small.  After the gravitational wave has traveled for a distance $D$,  two components of the wave with frequency differing by $\Delta \omega$ will arrive at times separated by
$$\Delta t=\frac{D\Omega^2}{2}\Delta\left(\frac{1}{\omega^2}\right)\;,$$ which even for $D$ of order $1/H_0$ and $\Delta\omega$ of order $\omega$ is  less than the period $2\pi/\omega$ of the oscillation by a factor of order $H_0/\omega$.   It appears that WIMPs can have no detectable effect on the gravitational waves observed from sources at moderate redshift.

\vspace{10pt}

\section{Primordial Gravitational Waves}

As a second application, we consider the effect of cold dark matter on primordial gravitational waves. In much of what follows we will consider WIMP dark matter for concreteness, but the discussion generalizes to more general models of dark matter. Let us begin by summarizing the key events during cosmic history that are important for our treatment of the effects of WIMP dark matter on primordial gravitational waves. At early times WIMPs are relativistic and are in thermal equilibrium with the particles of the standard model. As the universe cools, the dark matter particles become non-relativistic. Shortly after this time, when the temperature of the medium has dropped to $\approx 1/30$ of the WIMP mass, inelastic processes are no longer efficient enough to keep the dark matter particles in chemical equilibrium and the comoving number density of dark matter particles becomes constant. However, elastic scattering still occurs rapidly and keeps the WIMPs in kinematic equilibrium with the standard model. As the universe cools further, elastic scattering between the dark matter particles and standard model particles becomes inefficient as well, WIMPs kinetically decouple and become free-streaming. Astrophysical sources emit gravitational waves long after kinetic decoupling when the dark matter is already free-streaming. In contrast, depending on their frequency, primordial gravitational waves may propagate during earlier epochs when the dark matter was still in kinetic equilibrium or even relativistic.

We will refer to gravitational waves that enter the horizon after kinetic decoupling as long modes. For typical WIMPs, these have frequencies of at most a few times $10^{-12}$ Hz today, and can only be accessed through measurements of the polarization of the cosmic microwave background. We call modes that enter the horizon before kinetic decoupling but after the dark matter has become non-relativistic intermediate modes. These modes have frequencies between $10^{-12}$ and $\sim 10^{-5}$ Hz, and fall into the frequency range observable with pulsar timing arrays. Modes accessible with DECIGO~\cite{Kawamura:2008zza} or BBO~\cite{Phinney} enter the horizon when the dark matter particles are still relativistic, and we refer to them as short modes.

\paragraph{Long modes}\mbox{}\\
We first discuss effects on modes with wavelengths that can be accessed through measurements of the polarization of the cosmic microwave background. In linear perturbation theory primordial gravitational waves generate B-mode polarization whereas density perturbations do not. So the search for B-mode polarization of the CMB is an indirect search for gravitational waves. Lensing of the CMB by large scale structure between us and the surface of last scattering also generates B-mode polarization and in practice limits the multipoles for which we can extract information about primordial gravitational waves to less than a few hundred. 

The contribution to the CMB anisotropies at multipole $\ell$ is dominated by gravitational waves with wave number $k=a_L\ell/d_L$, where $a_L$ is the value of the scale factor at last scattering, and $d_L$ is the angular diameter distance to the surface of last scattering. For a flat geometry
\begin{equation}
d_L=\frac{1}{H_0(1+z_L)}\int_{1/(1+z_L)}^1\frac{dx}{\Omega_r+\Omega_m x+\Omega_\Lambda x^4}\approx 13\,{\rm Mpc}^{-1}\,.
\end{equation}
So the CMB allows us to access gravitational waves with comoving wave numbers $k\lesssim0.03\,{\rm Mpc}^{-1}$. These modes entered the horizon at a redshift of $z\lesssim 10^4$ long after kinetic decoupling of the dark matter. The anisotropic stress for the modes of interest is then well approximated by equation~(\ref{eq:piij}). Furthermore, by this time these modes have at most undergone a few oscillations so that the anisotropic stress for the modes accessible in the CMB further simplifies to~(\ref{eq:piij2}) and~(\ref{eq:E}).

In sections V and VI we found analytic solutions to the field equations in the presence of non-relativistic collisionless matter for wave frequencies much greater than the Hubble expansion rate, either using the WKB approximation to deal with general expansion rates, or in the special case of constant $a(t)$, where this approximation is unnecessary. We are now concerned with gravitational wave frequencies comparable to the expansion rate. Unfortunately there is no way to find analytic solutions of the field equations for Robertson-Walker scale factors $a(t)$ with arbitrary time-dependence. However, we can find solutions during the matter and radiation dominated eras most relevant to the CMB.

To treat the time evolution during the matter and radiation dominated eras, it is convenient to introduce the independent variable $y=a/a_{\rm eq}$, where $a_{\rm eq}$ is the scale factor at matter-radiation equality, and write equation~(\ref{eq:eom_hij}) as\footnote{This equation is valid after electrons and positrons have frozen out.}
\begin{equation}\label{eq:eom_hmr}
(1+y)\frac{d^2}{dy^2}h_{ij}(\mathbf{x},t)+\left(\frac{2}{y}+\frac{5}{2}\right)\frac{d}{dy}h_{ij}(\mathbf{x},t)+\varkappa^2 h_{ij}(\mathbf{x},t)=-\frac{4\epsilon}{y^3}(h_{ij}(\mathbf{x},t)-h_{ij}(\mathbf{x},t_1))\,,
\end{equation}
with $\epsilon=\mathcal{E}/a_{\rm eq}^5\rho_{\rm m\,eq}$ the fraction of the energy density of the dark matter particles stored in kinetic energy at matter-radiation equality, and $\varkappa=\sqrt{2}q/a_{\rm eq}H_{\rm eq}$. The solution to this equation cannot be written in closed form, but we can find solutions for $\varkappa\ll1$ and $\varkappa\gg1$. 

Let us first consider modes that enter the horizon after matter-radiation equality for which $\varkappa\ll1$. For modes outside the horizon at last scattering $h_{ij}(\mathbf{x},t)\approx h_{ij}(\mathbf{x},t_1)$ and the anisotropic stress vanishes. So we expect the evolution of the gravitational waves to be unaffected by the presence of cold dark matter. To be more quantitative, we can treat both the gradients and the anisotropic stress as a perturbation. Introducing the mode expansion
\begin{equation}\label{eq:h_mode}
h_{ij}(\mathbf{x},t)=\sum_{\lambda=\pm 2}\int\!d^3q\,\beta(\mathbf{q},\lambda)e_{ij}(\hat{q},\lambda)h_q(t)e^{i\mathbf{q}\cdot \mathbf{x}}\,,
\end{equation}
the general solution to the homogeneous equation is given by a linear combination of
\begin{equation}
h^1_q(y)=1\qquad{\rm and}\qquad h^2_q(y)=\left(\frac12\ln\frac{\sqrt{1+y}+1}{\sqrt{1+y}-1}-\frac{\sqrt{1+y}}{y}\right)\,.
\end{equation}
The second solution diverges like $1/y$ for small $y$ and it is the first the solution that is of interest in the context of primordial gravitational waves.
With help of the Green's function 
\begin{equation}
G(y,z)=\frac{z}{2y\sqrt{1+z}}\left(-2z\sqrt{1+y}+2y\sqrt{1+z}+yz+\ln\frac{\sqrt{1+y}+1}{\sqrt{1+y}-1}-\ln\frac{\sqrt{1+z}+1}{\sqrt{1+z}-1}\right)\theta(y-z)\,.
\end{equation}
we can write the solution at leading order in $\varkappa^2$ as
\begin{equation}\label{eq:hq0}
h_q^{(0)}(y)=h_q^{o}\left[1+\frac{2\varkappa^2}{15 y}\left(8-8\sqrt{1+y}-3y^2+4y\left(1+\ln\frac{y}{4}+\ln\frac{\sqrt{1+y}+1}{\sqrt{1+y}-1}\right)\right)\right]\,.
\end{equation}
The leading contribution from anisotropic stress also arises at order $\varkappa^2$ and is given by
\begin{equation}
h_q^{(1)}(y)=h_q^{(1)}(y_\star)-4\epsilon\int_{y_\star}^y dz G(y,z)\frac{h_q^{(0)}(z)-h_q^o}{z^3}\,,\label{eq:longGreen}
\end{equation}
where $y_\star$ is late enough for collisions to be negligible but early enough so the mode is far outside the horizon, and $h_q^{(1)}(y_\star)$ is the contribution generated by up to this point. We will compute it in section VII, for now we simply give the result
\begin{equation}
h_q^{(1)}(y_\star)=h_q^o\left(1+\frac{\epsilon\varkappa^2y_\star}{3}+C_\omega\right)\,,
\end{equation}
where $C_\omega$ is negative and describes a small amount of damping generated by collisions around the time of kinetic decoupling. It is of order $\epsilon\varkappa^2a_{\rm kd}/a_{\rm eq}$ and is suppressed relative to the terms of interest by $a_{\rm kd}/a_{\rm eq}\ll1$, where $a_{\rm kd}$ is the scale factor at kinetic decoupling, and we can safely neglect it.

The result cannot be written in closed form for general $y$ but becomes simple in the radiation and matter dominated epochs
\begin{eqnarray}
&&h_q(y)\to h_q^o\left(1-\frac16\varkappa^2y^2+\frac{\epsilon\varkappa^2 y}{3}\right) \hskip 1.8cm\qquad{\rm for}\qquad y\ll1\,,\label{eq:hqmatt}\\
&&h_q(y)\to h_q^o\left(1-\frac25\varkappa^2y+\frac{4\epsilon\varkappa^2(8\zeta(3)-7)}{15}\right)\hskip 1.5mm\qquad{\rm for}\qquad y\gg1\,.\label{eq:longmatt}
\end{eqnarray}
Since $8\zeta(3)-7\approx2.6>0$, we see that modes outside the horizon during last scattering receive a small scale-dependent boost. 
\begin{figure}[t]
\begin{center}
\includegraphics[width=4in]{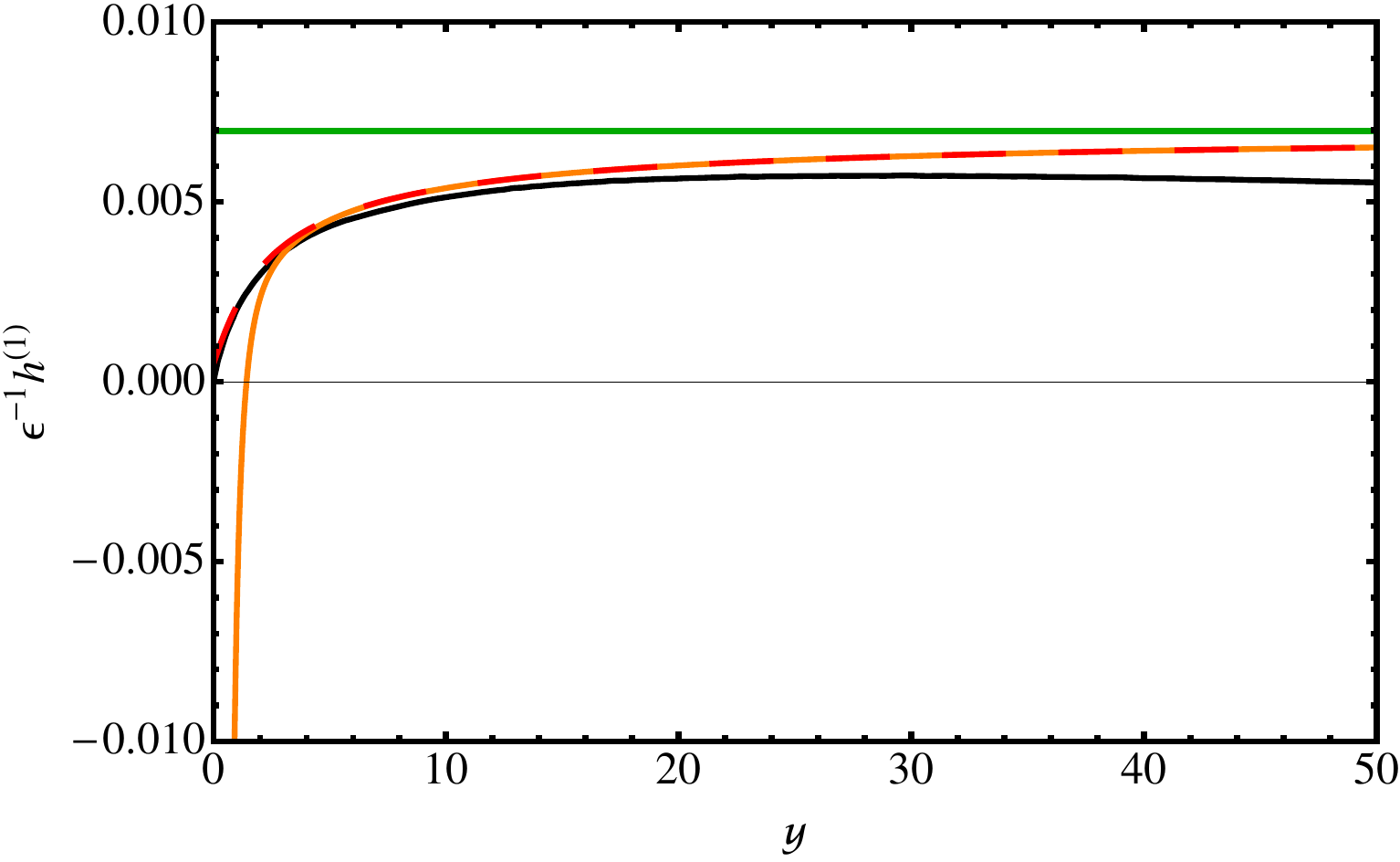}
\caption{The effect of collisionless matter on the time evolution of a mode with $\varkappa=1/10$. We show the limiting form given in equation~(\ref{eq:longmatt}) (green), the approximation given in equation~(\ref{eq:longapp}) (orange), the full expression based on equation~(\ref{eq:longGreen}) (dashed red), and the difference between the numerical solutions of the equation of motion with and without anisotropic stress (black). \label{fig:long}}
\end{center}
\end{figure}
Since last scattering occurs for $y\approx 3$, this simple limiting form does not capture the effect on the CMB accurately, but we can expand the result to higher orders, and find that the solution is given by $h_q(y)=h_q^{(0)}(y)+h_q^{(1)}(y)$ with $h_q^{(0)}(y)$ given by equation~(\ref{eq:hq0}) and the leading effect due to collisionless matter given by
\begin{eqnarray}
&&\hskip-1cm h_q^{(1)}(y)= 4\epsilon\varkappa^2 h_q^o\left(\frac{8\zeta(3)-7}{15}-\frac{4}{5y}+\frac{8 (15 + 2 \pi^2) }{135y^{3/2}}\right.\nonumber\\
&&\left.\hskip 2.05cm+\frac{4(7 + 2 \ln(y/4))}{15 y^2}-\frac{4 (15 + 2 \pi^2)}{225y^{5/2}}-\frac{32 (2 + \ln(y/4))}{135 y^3}\right)+\mathcal{O}(y^{-7/2})\label{eq:longapp}
\end{eqnarray}
The limiting form~(\ref{eq:longmatt}), the approximation~(\ref{eq:longapp}), and the result at order $\varkappa^2$ and linear in $\epsilon$ based on equation~(\ref{eq:longGreen}) are compared to the numerical result in Figure~\ref{fig:long} for $\varkappa=1/10$. The difference between the numerical result and our approximation for large $y$ arises because the mode is about to enter the horizon.

We see that the effect is highly suppressed and unobservably small for any upcoming or planned CMB experiment both because the fraction of the energy density stored in kinetic energy density of the dark matter is very small and because for these modes $\varkappa\ll1$.

Let us now turn to modes with $\varkappa\gg1$. These modes enter the horizon at a time when the energy density of the universe is dominated by radiation. To find their time evolution, we will first find the solution during radiation domination and then match it onto the WKB solution~(\ref{eq:WKBV}) to extend it to late times. 

\begin{figure}[t]
\begin{center}
\includegraphics[width=4in]{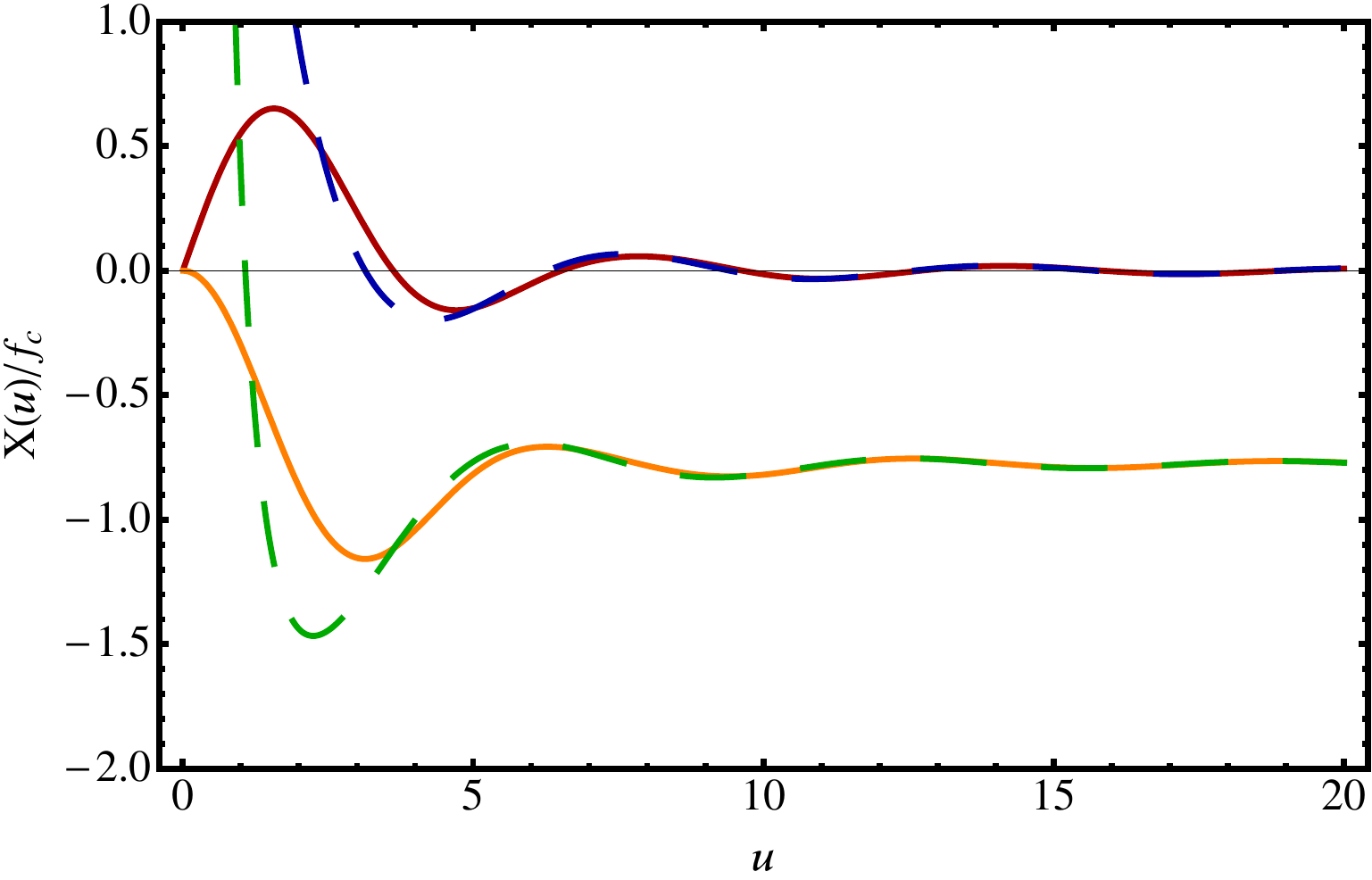}
\caption{$C(u)$ and $D(u)$ as defined in equations~(\ref{eq:Cnr}),~(\ref{eq:Dnr}). We show the exact results for $C(u)$ (orange) and $D(u)$ (red), and the limiting forms~(\ref{eq:Cnrinf}),~(\ref{eq:Dnrinf}) valid for $u\gg1$ for $C(u)$ (dashed blue) and $D(u)$ (dashed green). 
\label{fig:CDnr}}
\end{center}
\end{figure}

In the radiation dominated period, $y\ll1$, the equation of motion for gravitational waves~(\ref{eq:eom_hmr}) simplifies and the mode functions will only depend on $y$ through $u=\varkappa y$. It is then convenient to write the equation of motion as 
\begin{equation}
\frac{d^2}{du^2}h_q(u)+\frac{2}{u}\frac{d}{du}h_q(u)+h_q(u)=-\frac{4\epsilon\varkappa}{u^3}(h_q(u)-h_q(u_1))\,.
\end{equation}
The general solution of the homogeneous differential equation is a superposition of the solutions
\begin{eqnarray}
&&h_q^1(u)=\frac{\sin(u)}{u}\,,\label{eq:h1rad}\\
&&h_q^2(u)=\frac{\cos(u)}{u}\,.\label{eq:h2rad}
\end{eqnarray}
The second solution diverges for small $u$ and consequently decays outside the horizon so that the first solution is relevant for primordial gravitational waves. It is normalized so that $h^1_q(0)=1$.
In this case we can write the Green's function as
\begin{eqnarray}\label{eq:Greens_rad}
G(u,v)=\frac{v\sin(u-v)}{u}\theta(u-v)=v^2\left[h_q^1(u)h_q^2(v)-h_q^2(u)h_q^1(v)\right]\theta(u-v)\,,
\end{eqnarray}
and we can formally write the solution to the inhomogeneous equation as
\begin{equation}
h_q(u)=h_q^{(0)}(u)-4\epsilon\varkappa \int_{u_\star}^u dv\, G(u,v)\frac{h_q(v)-h_q(v_1)}{v^3}\,.
\end{equation}
The integral and its derivative vanish at $u_\star$ so the homogeneous solution must be chosen to satisfy the desired initial conditions. We can write it as
\begin{equation}
h^{(0)}_q(u)=Ah_q^1(u)+Bh_q^2(u)\,,\label{eq:hAB}
\end{equation}
with
\begin{eqnarray}
A&=&h_q(u_\star) \left(\cos(u_\star) + u_\star \sin(u_\star)\right)+h_q'(u_\star) u_\star \cos(u_\star) \,,\\
B&=&h_q(u_\star) \left(u_\star \cos(u_\star)-\sin(u_\star)\right) -h_q'(u_\star) u_\star \sin(u_\star)\,.
\end{eqnarray}
To first order in $\epsilon \varkappa$ we can write the solution as a superposition of the two solutions of the homogeneous solution, albeit with time dependent coefficients
\begin{eqnarray}
h_q(u)&=&A\left[(1+C(u))h_q^1(u)+ D(u)h_q^2(u)\right]\nonumber\\
&+& B\left[ E(u)h_q^1(u)+(1+F(u))h_q^2(u)\right]\,,
\end{eqnarray}
with
\begin{eqnarray}
&&C(u)= -4\epsilon\varkappa\int_{u_\star}^u \frac{dv}{v}\,h_q^2(v)\left((h_q^1(v)-h_q^1(v_1)\right)\,,\label{eq:Cnr}\\
&&D(u)= 4\epsilon\varkappa\int_{u_\star}^u \frac{dv}{v}\,h_q^1(v)\left(h_q^1(v)-h_q^1(v_1)\right)\,,\label{eq:Dnr}\\
&&E(u)= -4\epsilon\varkappa\int_{u_\star}^u \frac{dv}{v}\,h_q^2(v)\left(h_q^2(v)-h_q^2(v_1)\right)\,,\label{eq:Enr}\\
&&F(u)= 4\epsilon\varkappa\int_{u_\star}^u \frac{dv}{v}\,h_q^1(v)\left(h_q^2(v)-h_q^2(v_1)\right)\,.\label{eq:Fnr}
\end{eqnarray}
These integrals can all be expressed in terms of trigonometric functions, sine and cosine integrals, but we will not give the general formulae and work in various limits. For primordial gravitational waves we expect $h_q^{(0)}(u)=h_q^{o}h_q^1(u)$ so that
\begin{equation}
h_q(u)=h_q^o(1+C(u))h_q^1(u)+h_q^o D(u)h_q^2(u)\,,
\end{equation}
or
\begin{equation}
h_q^{(1)}(u)=h_q^oC(u)h_q^1(u)+h_q^o D(u)h_q^2(u)\,,
\end{equation}
and we only need the behavior of $C(u)$ and $D(u)$. As we will see, this is not entirely accurate because a small departure from $A=1$ and $B=0$ is generated around the time of kinetic decoupling, and as we will see
\begin{equation}
A=1+\epsilon\varkappa \frac{2u_\star}{3}+C_\omega\qquad\text{and}\qquad B=-\epsilon\varkappa \frac{u_\star^2}{3}\,.
\end{equation}
The amount of damping generated around kinetic decoupling, $C_\omega$, is calculated below. For now, it suffices to know that it is of order $\epsilon\varkappa^2a_{\rm kd}/a_{\rm eq}$, where $a_{\rm kd}$ is the scale factor at kinetic decoupling (defined more precisely below). Since $C_\omega$ is suppressed not only by $\epsilon$ but also by $a_{\rm kd}/a_{\rm eq}$ we can safely neglect it in our discussion here. This implies that we have
\begin{equation}
h_q^{(1)}(u)=h_q^o\left(C(u)+\epsilon\varkappa \frac{2u_\star}{3}\right)h_q^1(u)+h_q^o \left(D(u)-\epsilon\varkappa \frac{u_\star^2}{3}\right)h_q^2(u)\,,
\end{equation} 
For small $u$ it is easy to see that we can drop the additional terms provided we set $u_\star=0$ in equations~(\ref{eq:Cnr}) and~(\ref{eq:Dnr}), and we will do so in what follows. For modes that are far outside the horizon when the particles become non-relativistic $v_1\ll1$. The leading correction is quadratic in $v_1$, and we will take $v_1\to0$. We will need the limiting forms for $u\ll1$ and $u\gg1$. For small arguments we find
\begin{eqnarray}
&&C(u)\to \epsilon\varkappa \frac{2u}{3}+\mathcal{O}(u^3)\,,\\
&&D(u)\to -\epsilon\varkappa \frac{u^2}{3}+\mathcal{O}(u^4)\,,
\end{eqnarray}
whereas for large arguments
\begin{eqnarray}
&&C(u)\to 4\epsilon\varkappa\frac{\sin(u)}{u^2}+\mathcal{O}(1/u^3)\,,\label{eq:Cnrinf}\\
&&D(u)\to 2\epsilon\varkappa\left(1-2\ln 2+\frac{2\cos(u)-1/2}{u^2}\right)+\mathcal{O}(1/u^3)\label{eq:Dnrinf}\,.
\end{eqnarray}
This leads to a solution for the mode function far outside the horizon of
\begin{equation}
h_q(y)=h_q^o\left(1-\frac16\varkappa^2y^2+\frac{\epsilon\varkappa^2y}{3}\right)+\mathcal{O}(y^3)\,,
\end{equation}
in agreement with equation~(\ref{eq:hqmatt}). Once the mode is deep inside the horizon, it approaches
\begin{equation}\label{eq:radsol}
h_q(y)=h_q^o\left(\frac{\sin(\varkappa y)}{\varkappa y}-\frac{2\epsilon\varkappa \cos(\varkappa y)(2\ln 2-1)}{\varkappa y}\right)+\mathcal{O}(\varkappa^{-3}y^{-3})\,.
\end{equation}
We see that the dark matter has no effect on the amplitude (besides the small effect generated around kinetic decoupling we neglected) but introduces a small phase shift. Since we will need it later, let us also record its derivative
\begin{equation}\label{eq:dradsol}
h_q'(y)=\varkappa h_q^o\left(\frac{\cos(\varkappa y)}{\varkappa y}+\frac{2\epsilon\varkappa \sin(\varkappa y)(2\ln 2-1)}{\varkappa y}\right)+\mathcal{O}(\varkappa^{-2}y^{-3})\,.
\end{equation}
The behavior of the functions $C(u)$ and $D(u)$ and the comparison to the limiting forms~(\ref{eq:Cnrinf}),~(\ref{eq:Dnrinf}) are shown in Figure~\ref{fig:CDnr}.

This solution is valid deep inside the horizon and during the radiation dominated era. To find the solution at later times, we can match it to the WKB approximation we derived in section V. Equation~(\ref{eq:WKBV}) becomes
\begin{equation}\label{eq:WKBinh}
h_q(y)=h_q^1(y)\left[h_q(y_\star)+h_q(y_1)A(y)\vphantom{\varpi^{-1}y}\right]+h_q^2(y)\left[\varpi(y_\star)^{-1}h_q'(y_\star)+h_q(y_1)B(y)\right]\,,
\end{equation}
with
\begin{equation}
\varpi(y)=\frac{\sqrt{\varkappa^2+\frac{4\epsilon}{y^3}}}{\sqrt{1+y}}\,,
\end{equation}
the functions
\begin{eqnarray}
&&\hskip -.5cm h_q^1(y)=\frac{y_\star}{y}\cos\left(2\varkappa \left(\sqrt{1 + y} - \sqrt{1 + y_\star}\right)\right)-\frac{\epsilon}{\varkappa y y_\star}\sin\left(2\varkappa \left(\sqrt{1 + y} - \sqrt{1 + y_\star}\right)\right)\,,\\
&&\hskip -.5cm h_q^2(y)=\frac{y_\star}{y}\sin\left(2\varkappa \left(\sqrt{1 + y} - \sqrt{1 + y_\star}\right)\right)+\frac{\epsilon}{\varkappa y y_\star}\cos\left(2\varkappa \left(\sqrt{1 + y} - \sqrt{1 + y_\star}\right)\right)\,,
\end{eqnarray}
and to leading order in $\epsilon$ 
\begin{eqnarray}
&&A(y)= \frac{4 \epsilon(1+y) \cos(2 \varkappa (\sqrt{1 + y} - \sqrt{1 + y_\star}))}{\varkappa^2 y^2}-\frac{4 \epsilon(1+y_\star)}{\varkappa^2 y_\star^2}\,,\\
&&B(y)= \frac{4 \epsilon(1+y) \sin(2 \varkappa (\sqrt{1 + y} - \sqrt{1 + y_\star}))}{\varkappa^2 y^2}\,.
\end{eqnarray}
So to first order in $\epsilon$ and deep inside the horizon, we obtain the solution
\begin{eqnarray}
h_q(y)&=&h_q^o h_q^1(y)\left[\frac{\sin(\varkappa y_\star)}{\varkappa y_\star}+\frac{2\epsilon \cos(\varkappa y_\star)(1-2\ln 2)}{y_\star}-\frac{4 \epsilon(1+y_\star)}{\varkappa^2 y_\star^2}\right.\nonumber\\
&&\hskip 3.22cm\left.+\frac{4 \epsilon(1+y) \cos(2 \varkappa (\sqrt{1 + y} - \sqrt{1 + y_\star}))}{\varkappa^2 y^2}\right]\nonumber\\
&+&h_q^o h_q^2(y)\left[\frac{\cos(\varkappa y_\star)}{\varkappa y_\star}-\frac{2\epsilon \sin(\varkappa y_\star)(1-2\ln 2)}{y_\star}\right.\nonumber\\
&&\hskip 3.28cm\left.+\frac{4 \epsilon(1+y_\star) \sin(2 \varkappa (\sqrt{1 + y} - \sqrt{1 + y_\star}))}{\varkappa^2 y^2}\right]\,.
\end{eqnarray}
Working to leading order in $\epsilon$, the dependence on $y_\star$ disappears as it had to and the evolution inside the horizon valid during both radiation and matter dominated eras is given by
\begin{equation}\label{eq:shortapp}
h_q(y)=h_q^o\left[\frac{\sin\left(2\varkappa \left(\sqrt{1 + y} - 1\right)\right)}{\varkappa y}-\frac{2\epsilon\varkappa (2\ln 2-1) \cos\left(2\varkappa \left(\sqrt{1 + y} - 1\right)\right)}{\varkappa y}\right]\,.
\end{equation}
\begin{figure}[t]
\begin{center}
\includegraphics[width=4in]{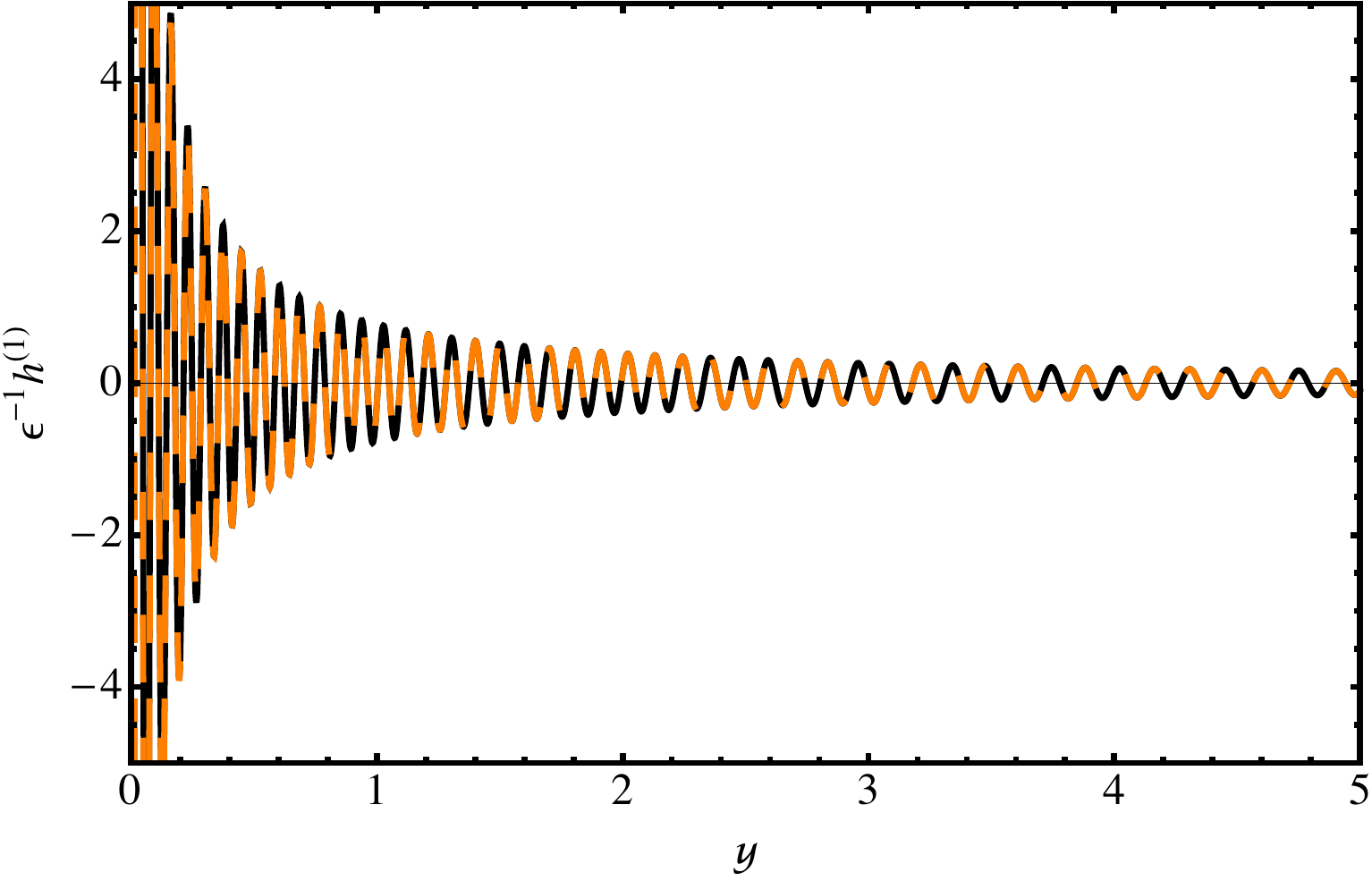}
\caption{The effect of collisionless matter on the time evolution of a mode with $\varkappa=100$. We show the term of order $\epsilon$ of the approximation to the mode function given in equation~(\ref{eq:shortapp}) (dashed orange) and the difference between the numerical solutions of the equation of motion with and without anisotropic stress (black). \label{fig:short}}
\end{center}
\end{figure}
We see that the gravitational waves acquire a small phase shift $\delta\varphi=-2\epsilon\varkappa(2\ln2-1)$. The analytic solution is compared to a numerical calculation in Figure~\ref{fig:short} for $\varkappa=100$. We see that the effect on modes that enter the horizon during the radiation dominated period is larger than the effect on modes that enter at later times, but since the fraction of the density in the kinetic energy of the dark matter is rather small, its effect on the degree scale polarization of the cosmic microwave background is also too small to be observed with upcoming or planned CMB experiments.

\paragraph{Intermediate modes}\mbox{}\\
We now turn to modes that enter the horizon when the dark matter is still in kinetic equilibrium but has already become non-relativistic. For a typical WIMP this corresponds to a gravitational wave frequency today below $\sim 10^{-5}$ Hz. 

As we briefly discussed after equation~(\ref{eq:Vlasov}), we expect collisions to be negligible if the collision term in the Boltzmann equation is much less than the transport term. The wavelength of the primordial gravitational waves, $\lambda$, redshifts like one power of the scale factor, the velocity of the dark matter particles, $v$ redshifts like $a^{-1}$ after and $a^{-1/2}$ before kinetic decoupling. The rate $\omega_{\rm r}$ at which energy is exchanged between standard model particles and the dark matter redshifts at least like $a^{-3}$ like the number density of standard model particles. So at late times when $\omega_{\rm r}\ll v/\lambda$ collisions are negligible, but they become important at early times. As a consequence we see that the anisotropic stress is no longer given by~(\ref{eq:piij}) and we will have to revisit the derivation in the presence of collisions. 

If the standard model particles interacting with the dark matter are much lighter than the dark matter particles, are relativistic and are in local thermal equilibrium, the Boltzmann equation becomes
\begin{eqnarray}\label{eq:boltzmann}
&&\hskip -0.5cm\frac{\partial n(\mathbf{p},\mathbf{x},t)}{\partial t}+\frac{p^i}{p^0}\frac{\partial}{\partial x^i}n(\mathbf{p},\mathbf{x},t)+\frac12\frac{\partial g_{kl}}{\partial x^i}\frac{p^k p^l}{p^0}\frac{\partial}{\partial p_i}n(\mathbf{p},\mathbf{x},t)=\nonumber\\
&&\hskip 3cm -2\langle\sigma v\rangle\Big[n(\mathbf{p},\mathbf{x},t)n(\mathbf{x},t)-n_{{\rm eq}}(p,\mathbf{x},t)n_{{\rm eq}}(\mathbf{x},t)\Big]\nonumber\\
&&\hskip 3cm+\omega_{\rm r}(t)\frac{\partial}{\partial p_i}\left[p_in(\mathbf{p},\mathbf{x},t)+g_{ij}(\mathbf{x},t)m T \frac{\partial}{\partial p_j} n(\mathbf{p},\mathbf{x},t)\right]\,,
\end{eqnarray}
where $T$ is the temperature of the standard model degrees of freedom, $\langle\sigma v\rangle$ is the thermally averaged dark matter annihilation cross section, $\omega_{\rm r}$ is the rate at which the standard model particles and dark matter particles exchange energies of order $kT$, and as before
\begin{equation}
p^i=g^{ij}(\mathbf{x},t)p_j\,,\qquad p^{0}=\sqrt{m^2+g^{ij}(\mathbf{x},t)p_ip_j}\approx m+\frac{g^{ij}(\mathbf{x},t)p_ip_j}{2m} \,,
\end{equation}
and
\begin{equation}
n(\mathbf{x},t)=\frac{1}{\sqrt{{\rm det}g(\mathbf{x},t)}}\int\! d^3p\, n(\mathbf{p},\mathbf{x},t)\,.
\end{equation}
In general, we expect the temperature to be a function of position and expect a small position dependent velocity of the medium, but because we are interested in tensor perturbations we will not need to include this.

In writing equation~(\ref{eq:boltzmann}), we have assumed that the dark matter only participates in interactions with the standard model particles, both in the form of the inelastic processes responsible for setting the freeze-out abundance, and in the form of the elastic processes required by crossing symmetry, but have neglected self-interactions. Of course, we only have very weak constraints on dark matter matter self-interactions, and these interactions may, in fact, well be significantly stronger than the interactions with the standard model that are included here, at least for some range of temperatures. However, we will see that our treatment of the effects of the minimal interactions that must be present for any WIMP included here will also allow us to understand the effects of self-interacting dark matter on gravitational waves.

Close to local thermal equilibrium the scattering rate is much higher than the rate of change in the temperature or the metric. We can thus neglect time derivatives acting on the metric or the temperature and see that the equilibrium distribution is 
\begin{equation}
n_{{\rm eq}}(\mathbf{p},\mathbf{x},t)=n_{{\rm eq}}\left(\frac{1}{2\pi m T}\right)^{3/2} \exp\left(-\frac{g^{ij}(\mathbf{x},t)p_ip_j}{2m T}\right)\,.
\end{equation}
Away from thermal equilibrium we should in general consider an Ansatz in which the temperature of the dark matter particles depends on position, but because we are interested in tensor perturbations we can consider an Ansatz in which it is only a function of time
\begin{equation}
n(\mathbf{p},\mathbf{x},t)=n(t)\left(\frac{1}{2\pi m T_{\rm dm}(t)}\right)^{3/2} \exp\left(-\frac{g^{ij}(\mathbf{x},t)p_ip_j}{2m T_{\rm dm}(t)}\right)+\delta n(\mathbf{p},\mathbf{x},t)\,.
\end{equation}
The first term on the right hand side is a solution to the Boltzmann equation in the absence of tensor perturbations provided the dark matter temperature and density obey
\begin{eqnarray}
\frac{1}{a^2}\frac{d}{dt}\left(a^2 T_{\rm dm}\right)&=& 2\omega_{\rm r}(t)(T-T_{\rm dm})\,,\label{eq:Tdmev}\\
\frac{1}{a^3}\frac{d}{dt}\left(a^3 n\right)&=&-2\langle\sigma v\rangle\left(n^2-n_{\rm eq}^2\right)\label{eq:nfreeze}\,.
\end{eqnarray}
So as expected $\delta n(\mathbf{p},\mathbf{x},t)$ is of first order in the metric perturbation, and as before we will write
\begin{equation}\label{eq:phasespace}
n(\mathbf{p},\mathbf{x},t)=\overline{n}(p)-\frac12h_{ij}(\mathbf{x},t)p_i\frac{\partial}{\partial p_j}\overline{n}(p)+\delta n(\mathbf{p},\mathbf{x},t)\,,
\end{equation}
with $\overline{n}(p,t)$ given by 
\begin{equation}\label{eq:nzero}
\overline{n}(p,t)=a^3n(t)\left(\frac{1}{2\pi m a^2 T_{\rm dm}}\right)^{3/2} \exp\left[{-\frac{p^2}{2m a^2T_{\rm dm}}}\right]\,.
\end{equation}
The equation for a plane wave, $\delta n(\mathbf{p},\mathbf{x},t)\propto\exp(i\mathbf{q}\cdot\mathbf{x})$, with wave vector $\mathbf{q}$ then becomes
\begin{eqnarray}\label{eq:linearBEs}
&&\hskip -1cm\frac{\partial \delta n(\mathbf{p},\mathbf{x},t)}{\partial t}+\frac{i \mathbf{p}\cdot \mathbf{q}}{a^2 m}\delta n(\mathbf{p},\mathbf{x},t)-\frac12\dot{h}_{ij}(\mathbf{x},t)\hat{p}_i\hat{p}_jp\frac{\partial}{\partial p}\overline{n}(p,t)=\nonumber\\
&&\hskip 1cm -2\omega_{\rm a}(t)\delta n(\mathbf{p},\mathbf{x},t)+\omega_{\rm r}(t)\frac{\partial}{\partial p_i}\left[p_i \delta n(\mathbf{p},\mathbf{x},t)+a^2 m T \frac{\partial}{\partial p_i} \delta n(\mathbf{p},\mathbf{x},t)\right]\,,
\end{eqnarray}
where we denoted the annihilation rate by
\begin{equation}
\omega_{\rm a}(t)=\langle\sigma v\rangle n(t)\,,
\end{equation}
and we have used 
\begin{equation}\label{eq:intdn}
\int d^3 p \,\delta n(\mathbf{p},\mathbf{x},t)=0\,,
\end{equation}
because gravitational waves do not generate fluctuations in the number density.\footnote{We will see a more rigorous justification for this below.}

Before we consider the general case, let us consider wavelengths for which the medium behaves like a viscous fluid. At leading non-trivial order in the derivative expansion, taking $H\ll \omega_{\rm r}$, $q/a\ll \omega_{\rm r}$ and using $\omega_{\rm a}\ll\omega_{\rm r}$, the perturbation to the phase space density must satisfy
\begin{equation}
\frac{\partial}{\partial p_i}\left[p_i \delta n(\mathbf{p},\mathbf{x},t)+a^2 m T \frac{\partial}{\partial p_i} \delta n(\mathbf{p},\mathbf{x},t)\right]=-\frac1{2\omega_{\rm r}}\dot{h}_{ij}(\mathbf{x},t)p_i\frac{\partial}{\partial p_j}\overline{n}(p,t)\,.
\end{equation}
Because $h_{ij}$ is traceless, we can commute $p_i$ with the derivative and the first integration is trivial. Remembering that the perturbation must vanish as the gravitational wave amplitude is taken to zero, we have
\begin{equation}
p_i \delta n(\mathbf{p},\mathbf{x},t)+a^2 m T \frac{\partial}{\partial p_i} \delta n(\mathbf{p},\mathbf{x},t)=-\frac{1}{2\omega_{\rm r}}\dot{h}_{ij}(\mathbf{x},t)p_j\overline{n}(p,t)\,.
\end{equation}
Since $\delta n(\mathbf{p},\mathbf{x},t)$ is a scalar that vanishes as the gravitational wave is taken to zero, and the metric perturbation is transverse and traceless we consider an Ansatz of the form
\begin{equation}
\delta n(\mathbf{p},\mathbf{x},t)=\dot{h}_{ij}(\mathbf{x},t)\hat{p}_i\hat{p}_j\tilde\Delta(q,p,t)\,.
\end{equation}
Introducing the shorthand notation $\dot{h}=\dot{h}_{kl}(\mathbf{x},t)\hat{p}_k\hat{p}_l$, the resulting equation is
\begin{eqnarray}
&&\hskip -1cm\hat{p}_i\dot{h}\tilde\Delta(q,p,t)+\frac{a^2m T}{p^2}\left[2\dot{h}_{ij}\hat{p}_j\tilde\Delta(q,p,t)+\hat{p}_i\dot{h}\left(-2\tilde\Delta(q,p,t)+p\frac{\partial}{\partial p}\tilde\Delta(q,p,t)\right)\right]\nonumber\\
&&\hskip 9cm=-\frac{1}{2\omega_{\rm r}}\dot{h}_{ij}\hat{p}_j\overline{n}(p,t)\,.
\end{eqnarray}
The coefficients of $\hat{p}_i$ and $\dot{h}_{ij}\hat{p}_j$ must vanish independently and from the term proportional to $\dot{h}_{ij}\hat{p}_j$ we can read off 
\begin{equation}\label{eq:soldn}
\tilde\Delta(q,p,t)=-\frac{p^2\overline{n}(p,t)}{4\omega_{\rm r} a^2m T}\,.
\end{equation}
Equation~(\ref{eq:Tdmev}) leads to
\begin{equation}\label{eq:TchiT}
T_{\rm dm}=T\left(1-\frac{H}{2\omega_{\rm r}}\right)\,,
\end{equation}
so that for $H\ll\omega_{\rm r}$ the dark matter temperature is well approximated by that of the standard model particles, $T_{\rm dm}\approx T$, and we see that the terms proportional to $\hat{p}_i$ also vanish for $\tilde\Delta(q,p,t)$ given by~(\ref{eq:soldn}). The perturbation to the phase space density in this approximation is then 
\begin{equation}\label{eq:dnvisc}
\delta n(\mathbf{p},\mathbf{x},t)=-\frac{p^2\overline{n}(p,t)}{4\omega_{\rm r} a^2m T}\dot{h}_{ij}(\mathbf{x},t)\hat{p}_i\hat{p}_j+\mathcal{O}\left(\frac{q^2}{a^2\omega_{\rm r}^2}\right)\,.
\end{equation}
Substituting back into the Boltzmann equation~(\ref{eq:linearBEs}), we see that the terms we are neglecting are indeed of order $q/a\omega_{\rm r}$ and $H/\omega_{\rm r}$ relative to the terms we are keeping. 
To compute the anisotropic stress, recall that the space-space components of the stress tensor are given by equation~(\ref{eq:Tijn}). For the Ansatz~(\ref{eq:phasespace}), the contribution linear in the metric perturbation simplifies to
\begin{eqnarray}\label{eq:Tij}
{\delta T^i}_j(\mathbf{x},t)&=&\frac{1}{a^5}\int d^3p\,\delta n(\mathbf{p},\mathbf{x},t)\frac{p_i p_j}{m}\,,
\end{eqnarray}
and the anisotropic stress is simply the transverse traceless part of this expression. We can perform the angular integrals with the identity
\begin{equation}
\int\frac{d^2\hat{p}}{4\pi}\hat{p}_i\hat{p}_j\hat{p}_k \hat{p}_l=\frac{1}{15}\Big[\delta_{ij}\delta_{kl}+\delta_{ik}\delta_{jl}+\delta_{il}\delta_{jk}\Big]\,,
\end{equation}
and the integral over the magnitude by recalling 
\begin{equation}
\frac{\partial}{\partial p}\overline{n}(p,t)=-\frac{p}{a^2m T_{\rm dm}}\overline{n}(p,t)\,,
\end{equation}
integrating by parts and using the definition of comoving kinetic energy density of the dark matter particles
\begin{equation}
\mathcal{E}(t)=\int\!d^3p\,\frac{p^2}{2m}\overline{n}(p,t)=\frac{3}{2}a^5 n T_{\rm dm}\approx \frac{3}{2}a^5 n T\,.
\end{equation}
This leads us to the anisotropic stress
\begin{equation}\label{eq:anisoE}
\pi_{ij}(\mathbf{x},t)=-\frac{\mathcal{E}(t) }{3a^5\omega_{\rm r}(t)}\dot{h}_{ij}(\mathbf{x},t)=-\frac{n T }{2\omega_{\rm r}(t)}\dot{h}_{ij}(\mathbf{x},t)\,,
\end{equation}
with the number density $n$ set by the usual freeze-out calculation. The equation of motion for gravitational waves before then simply becomes
\begin{equation}
\ddot{h}_q(t)+(3H(t)+\Gamma)\dot{h}_q(t)+\frac{q^2}{a^2(t)}h_q(t)=0\qquad{\rm with}\qquad\Gamma=8\pi G\frac{ n T}{\omega_{\rm r}}\,,
\end{equation}
so that the presence of the dark matter leads to some amount of damping of the gravitational waves. Repeating the above computation for a velocity gradient, we find that the shear viscosity of the medium is given by $\eta=n T/2\omega_{\rm r}$, so that the damping rate is given by $\Gamma=16\pi G\eta$ consistent with \cite{Hawking:1966qi}. However, because
\begin{equation}
\frac{\mathcal{E}(t) }{3a^5\omega_{\rm r}(t)M_{\rm p}^2 }\ll \frac{H}{\omega_{\rm r}}H\ll H\,,
\end{equation}
the Hubble rate during this epoch is orders of magnitude larger than $\Gamma$. 
The effect is highly suppressed both because the energy density in dark matter particles is a subdominant contribution to the total energy density during radiation domination, and because $H\ll \omega_{\rm r}$ before kinetic decoupling. 

We know that $\omega_{\rm r}\approx H$ during kinetic decoupling so that the approximation does not allow us to follow modes through kinetic decoupling, and we can only use it to study the behavior of modes before kinetic decoupling while $q/a\omega_{\rm r}\ll1$ and $H/\omega_{\rm r}\ll1$. To follow modes through decoupling, we return to equation~(\ref{eq:linearBEs}) and rewrite it as a hierarchy of coupled ordinary differential equations. 
Recalling the mode expansion~(\ref{eq:h_mode}), we see that the equation only depends on the direction of the momentum of the dark matter particles through $\mu=\hat{p}\cdot\hat{q}$ and $e_{ij}(\hat{q},\lambda)\hat{p}^i\hat{p}^j$. In general, additional directional dependence could arise from the initial conditions, but we are interested in isotropic initial conditions so that the perturbation to the phase space density must be of the form
\begin{equation}\label{eq:dnchiansatz}
\delta n(\mathbf{p},\mathbf{x},t)=\sum_{\lambda=\pm 2}\int\!d^3q\,\beta(\mathbf{q},\lambda)e_{kl}(\hat{q},\lambda)\hat{p}_k\hat{p}_l \tilde\Delta(q,p,\mu,t) e^{i\mathbf{q}\cdot \mathbf{x}}\,.
\end{equation}
Given that the polarization tensor is transverse and traceless, we see that this Ansatz justifies equation~(\ref{eq:intdn}). As we show in Appendix A, expanding the perturbation to the phase space density in terms of orthonormal polynomials  
\begin{eqnarray}
&&\hskip -1cm\delta n(\mathbf{p},\mathbf{x},t)=\sum_{\lambda=\pm 2}\int\!d^3q\,\beta(\mathbf{q},\lambda)e_{ij}(\hat{q},\lambda)\hat{p}_i\hat{p}_je^{i\mathbf{q}\cdot \mathbf{x}}\nonumber\\ 
&&\hskip 2cm\times\sum_{\substack{\ell=2\dots\infty\\n=0\dots\infty}}(-i)^\ell(2\ell+1)\Delta_{n\,\ell}(q,t)\mathcal{L}_{n\,\ell}(z)\mathcal{P}_\ell(\mu)p\frac{\partial}{\partial p}\overline{n}(p,t)\label{eq:dnexp} \,,
\end{eqnarray}
where
\begin{equation}\label{eq:z}
\mathcal{P}_\ell(\mu)=\frac{P_\ell^2(\mu)}{1-\mu^2}\qquad{\rm and}\qquad\mathcal{L}_{n\,\ell}(z)=z^{\ell/2-1}L_n^{\ell+1/2}(z)\qquad{\rm where}\qquad z=\frac{p^2}{2a^2m T_{\rm dm}}\,,
\end{equation}
$L_n^k$ are generalized Laguerre polynomials and $P_\ell^m$ are associated Legendre polynomials, allows us to diagonalize the collision term and leads us to the Boltzmann hierarchy
\begin{eqnarray}\label{eq:Dnann}
&&\hskip -1.72cm\dot{\Delta}_{n\,\ell}(q,t)+\frac{q}{(2\ell+1)a}\left(\frac{2T_{\rm dm}} {m}\right)^{1/2}\Bigg[(\ell+2)\left(n+\ell+\frac32\right)\Delta_{n\,\ell+1}(q,t)\nonumber\\*
&&-n(\ell+2)\Delta_{n-1\,\ell+1}(q,t)+(\ell-2)\Delta_{n+1\,\ell-1}(q,t)-(\ell-2)\Delta_{n\,\ell-1}(q,t)\Bigg]=\nonumber\\*
&&\hskip 1cm-\frac{1}{30}\dot{h}_q(t)\delta_{\ell2}\delta_{n0}-(2n+\ell)\omega_{\rm r}(t)\frac{T}{T_{\rm dm}}\Delta_{n\,\ell}(q,t)-2\omega_{\rm a}(t)\frac{n_{{\rm eq}}^2}{n^2}\Delta_{n\,\ell}(q,t)\,,
\end{eqnarray}
and the anisotropic stress
\begin{equation}\label{eq:piD02}
\pi_q(t)=30n(t) T_{\rm dm}(t)\Delta_{02}(q,t)\,.
\end{equation}

We see that for non-relativistic dark matter particles the collision term is dominated by the elastic scattering processes as expected. Dark matter self-interactions introduce another source of damping on the right hand side. Assuming they are generated by an operator with comparable coefficient to that responsible for the interactions between the dark matter and the standard model, their effect would be suppressed just like that of annihilations because the dark matter is non-relativistic and its number density is small compared to that of light standard model degrees of freedom.

To find the initial conditions for equation~(\ref{eq:Dnann}), let us consider the system at a time when scattering is efficient and $q/a\omega_{\rm r}\ll1$ and $H/\omega_{\rm r}\ll1$. We see that in this limit all modes but the mode with $n=0$ and $\ell=2$ are rapidly driven to zero. Recalling that in this limit $T_{\rm dm}\approx T$, we find
\begin{eqnarray}
\Delta_{02}(q,t)&\to&-\frac{\dot{h}_q(t)}{60\omega_{\rm r}(t)}\,,\label{eq:limD0}\\
\Delta_{n\,\ell}(q,t)&\to&0\qquad{\rm for\; all\; others.}\label{eq:limDn}
\end{eqnarray}
The expansion~(\ref{eq:dnexp}) together with $\mathcal{L}_{02}(z)=1$ and $\mathcal{P}_2(\mu)=3$ then implies
\begin{equation}
\delta n(\mathbf{p},\mathbf{x},t)=-\frac{p^2 \overline{n}(p,t)}{4\omega_{\rm r}a^2m T}\dot{h}_{ij}(\mathbf{x},t)\hat{p}_i\hat{p}_j\,,
\end{equation}
in agreement with our earlier result~(\ref{eq:dnvisc}). As a further consistency check consider gravitational wave emission at some time $t_1$ long after decoupling. Provided we are interested in the anisotropic stress at a time $t$ that is not too long after emission so that we still have
\begin{equation}
\int_{t_1}^t\!dt'\,\frac{q}{a(t')}\left(\frac{2T_{\rm dm}(t')}{m}\right)^{3/2}\ll1\,,
\end{equation}
all couplings between modes are negligible and we simply have
\begin{equation}
\Delta_{02}(q,t)=-\frac{1}{30}\left(h_q(t)-h_q(t_1)\right)\,,
\end{equation}
so that
\begin{equation}
\pi_{ij}(\mathbf{x},t)=-{n(t) T_{\rm dm}(t)}\left(h_{ij}(\mathbf{x},t)-h_{ij}(\mathbf{x},t_1)\right)\,,
\end{equation}
consistent with equation~(\ref{eq:piij2}) in section IV since the comoving kinetic energy density is given by $\mathcal{E}=3a^5n T_{\rm dm}/2$.

As long as the particles move a distance that is short compared to the wavelength of the gravitational wave on the time scale on which the dark matter and the standard model exchange energy, we have $(q/a)v\ll \omega_{\rm r}$ so that the higher multipole moments are driven to zero and the hierarchy reduces to
\begin{eqnarray}\label{eq:D0}
&&\hskip -1.72cm\dot{\Delta}_{02}(q,t)+2\omega_{\rm r}(t)\frac{T(t)}{T_{\rm dm}(t)}\Delta_{02}(q,t)=-\frac{1}{30}\dot{h}_q(t)\,.
\end{eqnarray}
All that remains is to find the initial conditions, but provided $q/a\omega_{\rm r}\ll 1$ around the time of freeze-out when $\omega_{\rm r}\gg H$, we know that the initial conditions are given by equation~(\ref{eq:limD0}), and the solution is
\begin{eqnarray}\label{eq:D02an}
\Delta_{02}(q,t)&=&-\frac{\dot{h}_q(t_1)}{60\omega_{\rm r}(t_1)}\exp\left[-2\int_{t_1}^t\!dt'\omega_{\rm r}(t')\frac{T(t')}{T_{\rm dm}(t')}\right]\nonumber\\*
&&-\frac{1}{30}\int_{t_1}^t\!dt'\,\dot{h}_q(t')\exp\left[-2\int_{t'}^t\!dt''\omega_{\rm r}(t'')\frac{T(t'')}{T_{\rm dm}(t'')}\right]\,.
\end{eqnarray}
Intermediate modes enter the horizon when the dark matter is non-relativistic, and we can take $t_1$ early enough so the mode is outside the horizon. In this case we can neglect the first term on the right hand side so that the time evolution for gravitational waves is governed by
\begin{equation}\label{eq:heomkd}
\ddot{h}_q(t)+3H\dot{h}_q(t)+\frac{q^2}{a^2}h_q(t)= -16\pi G n T_{\rm dm}\int_{t_1}^t\!dt'\,\dot{h}_q(t')\exp\left[-2\int_{t'}^t\!dt''\omega_{\rm r}(t'')\frac{T(t'')}{T_{\rm dm}(t'')}\right]\,.
\end{equation}

For modes that enter the horizon after kinetic decoupling the argument of the exponential is small and as expected the equation reduces to that studied in section IV. 

As an additional check, let us also consider modes that enter the horizon before kinetic decoupling when $\omega_{\rm r}\gg H$. For modes whose wave numbers satisfy $q/a\ll\omega_{\rm r}$, we see that the integral is dominated by times $t'$ that differ from $t$ by $\sim 1/\omega_{\rm r}$. Since the mode function varies on much longer time scales set by $q/a$ and $H$, we can approximate its argument by $t'\approx t$ and recover an anisotropic stress consistent with equation~(\ref{eq:piD02}) with $\Delta_{02}$ given by equation~(\ref{eq:limD0}). As we saw, this leads to an additional friction term, but the effect is much too small to be observable. 

As the universe expands, the rate $\omega_{\rm r}$ eventually drops below $q/a$. For modes that entered significantly before kinetic decoupling this happens while $\omega_{\rm r}\gg H$ so that $q/a\gg\omega_{\rm r}\gg H$. At this time $T_{\rm dm}\approx T $ and we can write the anisotropic stress as
\begin{eqnarray}
\pi_q(t)&=&n T \int_{t_1}^{t}\!dt'\,\dot{h}_q(t')\exp\left[-2\int_{t'}^t\!dt''\omega_{\rm r}(t'')\right]\,.
\end{eqnarray}
We can break up the integral into a contribution from the initial time $t_1$ to some time $t_\star$ when $q/a\gg \omega_{\rm r}\gg H$ and a contribution from $t_\star$ to the time of interest $t$
\begin{eqnarray}
\pi_q(t)&=& n T \int_{t_1}^{t_\star}\!dt'\,\dot{h}_q(t')\exp\left[-2\int_{t'}^{t_\star}\!dt''\omega_{\rm r}(t'')\right]\exp\left[-2\int_{t_\star}^t\!dt''\omega_{\rm r}(t'')\right]\nonumber\\
&&+n T \int_{t_\star}^{t}\!dt'\,\dot{h}_q(t')\exp\left[-2\int_{t'}^t\!dt''\omega_{\rm r}(t'')\right]\,,
\end{eqnarray}
The first term on the right hand side is then exponentially suppressed by the last factor provided $t$ is  at least a few $1/\omega_{\rm r}$ after $t_\star$, 
 and we can use the same trick as in equation~(\ref{eq:ibp}) to perform the integral on the second line because $q/a\gg \omega_{\rm r}\gg H$ for all $t'$. The equation of motion of the gravitational waves is then
\begin{equation}
\ddot{h}_q(t)+3H\dot{h}_q(t)+\frac{q^2}{a^2}h_q(t)= -16\pi G n T\left(h_q(t)-h_q(t_\star) \exp\left[-2\int_{t_\star}^t\!dt''\omega_{\rm r}(t'')\right]\right)\,.
\end{equation}
As long as $\omega_{\rm r}\gg H$, collisions rapidly erase the second term on the right hand side and the equation simplifies to the homogeneous equation
\begin{equation}\label{eq:heom_hom}
\ddot{h}_q(t)+3H\dot{h}_q(t)+\omega^2h_q(t)=0\qquad{\rm with}\qquad \omega^2=\frac{q^2}{a^2}+\frac{32\pi G \mathcal{E}}{3}\,,
\end{equation}
where $\mathcal{E}$ is the proper density of kinetic energy $\mathcal{E}=3n T/2$. So once $q/a\gg \omega_{\rm r}$, the only effect is the modified dispersion relation. We can then compute the phase shift caused by this modification throughout cosmic history as
\begin{equation}\label{eq:dphi}
\Delta\varphi=\int_{t_{\rm kd}}^{t_0}dt\frac{16\pi G\mathcal{E}}{3q/a(t)}\ll\frac{a_0H_0}{q}\ll 1\,,
\end{equation}
where $t_0$ denotes the present time. We see that even for primordial gravitational waves that entered the horizon before kinetic decoupling the modification to the dispersion relation has no observable effect. 

From this discussion, we see that modes are not significantly affected either at early times when $q/a\ll \omega_{\rm r}$ or once $q/a\gg \omega_{\rm r}$. What remains is to compute the effect of collisions around the time when $q/a\approx \omega_{\rm r}$. For this purpose it is convenient to introduce the independent variable $x=a/a_{\rm kd}$ and to define the Hubble rate at kinetic decoupling such that $H_{\rm kd}\equiv H(t_{\rm kd})= 2\omega_{\rm r}(t_{\rm kd})$. 
In this case equation~(\ref{eq:heomkd}) becomes
\begin{equation}
h_q''(x)+\frac{2}{x}h_q'(x)+\kappa^2 h_q(x)=-\frac{6 nT_{\rm dm} x^2}{\rho_{\rm kd}}\int_{x_1}^{x} dz h_q'(z)\exp\left[-\int_z^x dz' z' \hat{\omega}(z')\right]\,,
\end{equation}
where \mbox{$\hat{\omega}\left(y(t)\right)=\omega_{\rm r}(t)/\omega_{\rm r}(t_{\rm kd})$}, $\kappa=q/a_{\rm kd}H_{\rm kd}$ and $\rho_{\rm kd}$ is the energy density when $H(t_{\rm kd})= 2\omega_{\rm r}(t_{\rm kd})$. This equation neglects the effect introduced by the change in the number of relativistic degrees of freedom on the expansion rate studied in \cite{Watanabe:2006qe} because we are interested in small corrections introduced to the standard calculation of the gravitational wave spectrum by the velocity dispersion of the dark matter particles. We have set $T=T_{\rm dm}$ in the exponential because as we will see the effect of collisions on modes that enter before kinetic decoupling are most significant around the time when the wave number of the gravitational wave is comparable to $\omega_{\rm r}$, which occurs before kinetic decoupling when $T\approx T_{\rm dm}$. The integral on the right hand side receives negligible contributions at early times when the modes are frozen and we can set $x_1=0$.

We will keep $\hat\omega(y)$ general for now, but it may be helpful to know what behavior we expect. If the interactions between the dark matter particles and the standard model are controlled by a single operator, the dark matter is non-relativistic and the standard model particles are relativistic, the rate scales like $\omega_{\rm r}\propto T^{4+\beta}$. The value of $\beta$ is determined by the form of the interactions between dark matter and the standard model. An interaction between a non-relativistic scalar or fermionic dark matter particle and a relativistic scalar through a dimension four and five operator, respectively, would correspond to $\beta=0$, $\beta=2$ would describe a non-relativistic, fermionic dark matter particle interacting with a relativistic fermion through a dimension six operator, etc. 

The anisotropic stress is proportional to the fraction of the energy density stored in kinetic energy of the dark matter particles, which is small both because the dark matter particles are non-relativistic at the time of interest and because the universe is radiation dominated, justifying a perturbative treatment. Using the mode functions~(\ref{eq:h1rad}),~(\ref{eq:h2rad}), and the Green's function~(\ref{eq:Greens_rad}), the leading order solution is given by
\begin{equation}
h_q(x)=h_q^o(1+C(x))h_q^1(x)+h_q^o D(x)h_q^2(x)\,,
\end{equation}
with the functions 
\begin{eqnarray}
&&C(x)= -\int_{x_1}^x dy\,\kappa h_q^2(y)\frac{6nT_{\rm dm}y^4}{\rho_{\rm kd}}\int_{0}^y dz h_q^{1\prime}(z)\exp\left[-\int_z^y dz' z' \hat{\omega}(z')\right]\,,\label{eq:Ckd}\\
&&D(x)=\hphantom{-} \int_{x_1}^x dy\,\kappa h_q^1(y)\frac{6nT_{\rm dm}y^4}{\rho_{\rm kd}}\int_{0}^y dz h_q^{1\prime}(z)\exp\left[-\int_z^y dz' z' \hat{\omega}(z')\right]\,.\label{eq:Dkd}
\end{eqnarray}

Introducing the dark matter kinetic energy density at kinetic decoupling $\mathcal{E}_{\rm kd}$ and recalling that the temperature of the dark matter particles obeys equation~(\ref{eq:Tdmev}), we find
\begin{eqnarray}
&&C(x)= -\frac{4\mathcal{E}_{\rm kd}\kappa}{\rho_{\rm kd}}\int_{x_1}^x dy\, y \tau_{\rm dm} (y)h_q^2(y)\int_{0}^y dz h_q^{1\prime}(z)\exp\left[-\int_z^y dz' z' \hat{\omega}(z')\right]\,,\label{eq:Ckd2}\\
&&D(x)= \hphantom{-}\frac{4\mathcal{E}_{\rm kd}\kappa}{\rho_{\rm kd}}\int_{x_1}^x dy\, y\tau_{\rm dm}(y) h_q^1(y)\int_{0}^y dz h_q^{1\prime}(z)\exp\left[-\int_z^y dz' z' \hat{\omega}(z')\right]\,,\label{eq:Dkd2}
\end{eqnarray}
where $\tau_{\rm dm}=T_{\rm dm}/T_{\rm kd}$ is the solution of the differential equation
\begin{equation}
\frac{1}{y^3}\frac{d}{dy}\left(y^2\tau_{\rm dm}(y)\right)=\hat{\omega}\left(\frac{1}{y}-\tau_{\rm dm}\right)\label{eq:taudm}\,,
\end{equation}
that approaches $\tau_{\rm dm}(y)\to y^{-1}$ before kinetic decoupling when $y\ll1$. After kinetic decoupling the right hand side of the equation is negligible and the temperature of the dark matter particles redshifts like $y^{-2}$. Notice that here $T_{\rm kd}$ is the temperature of the standard model particles at kinetic decoupling so that $\mathcal{E}_{\rm kd}\equiv n(t_{\rm kd})T_{\rm kd}$ differs from the kinetic energy density in the dark matter particles at decoupling by a factor $\tau_{\rm dm}(1)$.

We can think of $C(x)$ as a change to the amplitude of the mode caused by collisions whereas $D(x)$ corresponds to the phase shift generated by them. Writing $\mathcal{E}_{\rm kd}=3\rho_{\rm m,\, kd}v_{\rm kd}^2/2$, we see that the effect is suppressed both because the velocity at decoupling for cold dark matter is of order $10^{-2}-10^{-3}$ and because decoupling typically happens deep in the radiation dominated era so that $\rho_{\rm m,\, kd}\ll \rho_{\rm kd}$.

For modes that enter the horizon long before kinetic decoupling $\kappa\gg 1$, and we already know from our earlier discussion that $C$ and $D$ do not receive significant contributions from very early or late times and we are interested in their behavior when $(q/a)\approx \omega_{\rm r}$ or $y\hat{\omega}\approx\kappa$ when $y\hat{\omega}\gg 1$ and $y\ll1$. In this case the integral over $z$ is dominated by $z\approx y$. Provided $(y\hat\omega)'\ll(y\hat\omega)^2$ we can change variables to $z=y+u$ and approximate the integral by expanding the argument of the exponential to leading order in $u$
\begin{equation}
\int_{0}^y dz h_q^{1\prime}(z)\exp\left[-\int_z^y dz' z' \hat{\omega}(z')\right]\approx \int_{-\infty}^0 du h_q^{1\prime}(y+u)\exp\left[u y\hat\omega(y)\right]\,.
\end{equation}
Expanding everywhere but in the trigonometric functions in $h_q^{1\prime}(z)$ to leading order in $u$ this leads to the following expression for $\kappa\gg1$
\begin{equation}\label{eq:intappr}
\int_{0}^y dz h_q^{1\prime}(z)\exp\left[-\int_z^y dz' z' \hat{\omega}(z')\right]\approx \frac{\kappa \left(1+y^2\hat{\omega}\right)\cos(\kappa y)+y(\kappa^2-\hat{\omega})\sin(\kappa y)}{\kappa y^2(\kappa^2+y^2\hat{\omega}^2)}\,.
\end{equation}
For large enough $y$ an additional constant contribution arises from a saddle point. However, this contribution decays rapidly for large $\kappa$ and in any case does not contribute once integrated against the oscillatory mode functions. So we will ignore it and work with~(\ref{eq:intappr}). 
\begin{figure}[t]
\begin{center}
\includegraphics[trim= 0cm 0cm 0cm 0.2cm, width=2.79in]{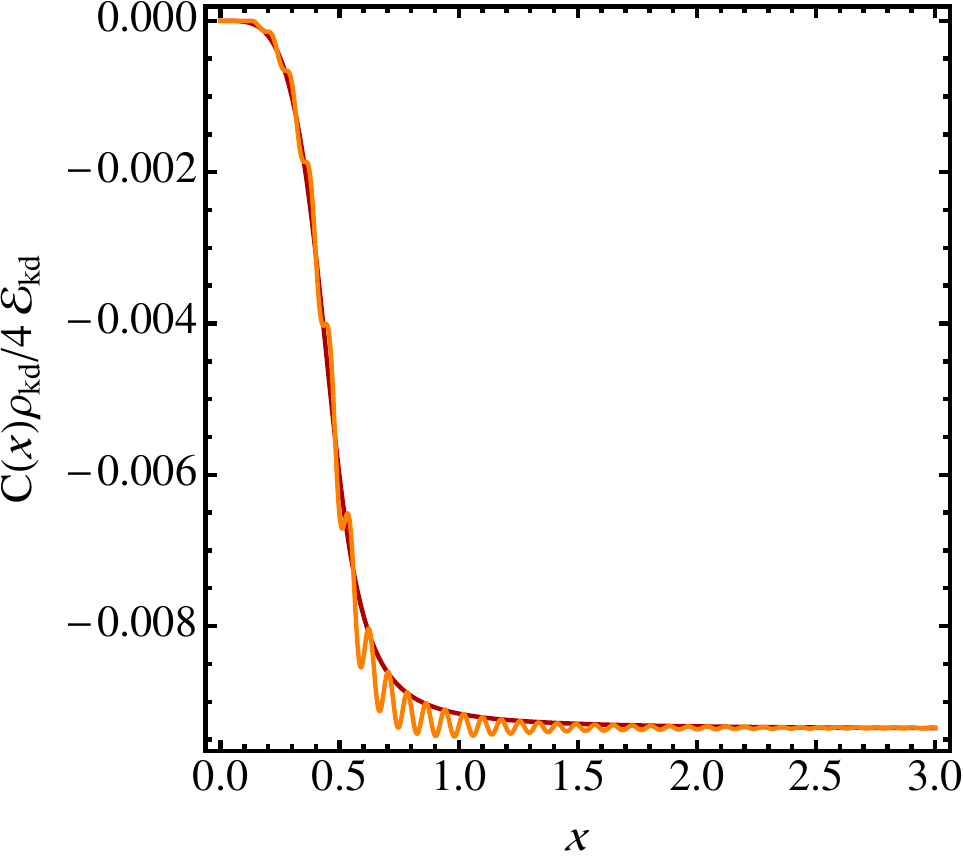}\hskip 0.5cm
\includegraphics[width=2.7in]{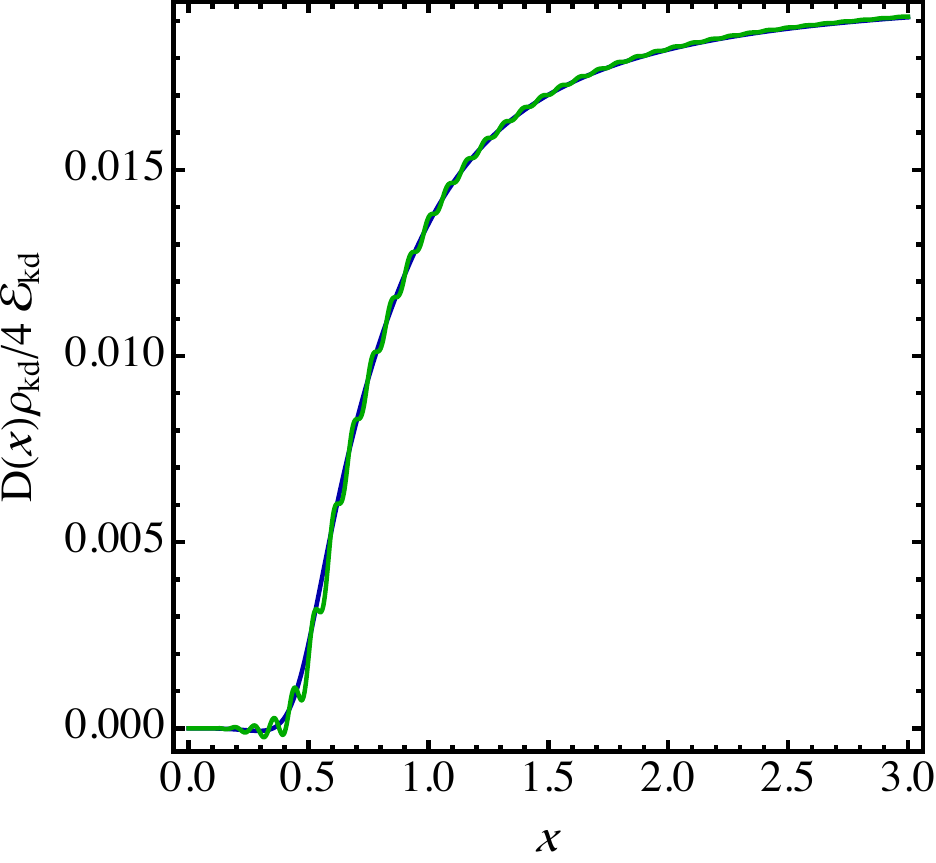}
\caption{Left: Comparison of equation~(\ref{eq:Cxan}) (red) with the results of a numerical calculation (orange). Right: Comparison of equation~(\ref{eq:Dxan}) (blue) with the results of a numerical calculation (green). The small oscillatory contributions were neglected in the analytic calculation because we were interested in the asymptotic behavior.\label{fig:CDkd}}
\end{center}
\end{figure}

Given equation~(\ref{eq:intappr}) we can easily find the dominant contributions to $C(x)$ and $D(x)$. Neglecting the suppressed oscillatory contributions, before kinetic decoupling when $\tau_{\rm dm}\approx y^{-1}$, we find 
\begin{eqnarray}
C(x)&\approx&-\frac{4\mathcal{E}_{\rm kd}}{\rho_{\rm kd}}\int_0^x dy\frac{1+y^2\hat\omega^2}{2y^3\left(\kappa^2+y^2\hat\omega^2\right)}\,,\label{eq:Cxan}\\
D(x)&\approx&\frac{4\mathcal{E}_{\rm kd}}{\rho_{\rm kd}}\int_0^x dy\frac{\kappa^2-\hat\omega}{2\kappa y^2\left(\kappa^2+y^2\hat\omega^2\right)}\label{eq:Dxan}\,.
\end{eqnarray}
As expected, the dominant contribution to the integrals arises when $\kappa\approx y\hat\omega$ or equivalently $q/a\approx\omega_{\rm r}$. 

Let us first consider the phase shift. Provided $\hat\omega$ decays more rapidly than $y^{-1}$, the phase shift at late times, when $\kappa y\gg1$ behaves like
\begin{equation}
D(x)\approx\frac{4\mathcal{E}_{\rm kd}}{\rho_{\rm kd}}\int^x dy\frac{1}{2\kappa y^2}\,,
\end{equation}
independent of the detailed behavior of $\hat\omega$ and consistent with the definition of the phase shift in equation~(\ref{eq:dphi}) valid for $q/a\gg\omega_{\rm r}$. 

Turning to the effect on the amplitude, the sign of $C(x)$ is negative so that gravitational waves are damped around the time when $q/a\approx\omega_{\rm r}$ as expected. 
 We show a comparison of a numerical computation with these results in Figure~\ref{fig:CDkd} for a representative wave number of $\kappa=40$ and for a rate that scales like a power law $\hat\omega(y)= y^{-(4+\beta)}$ with $\beta=2$. 

\begin{figure}[t]
\begin{center}
\includegraphics[width=4.6in]{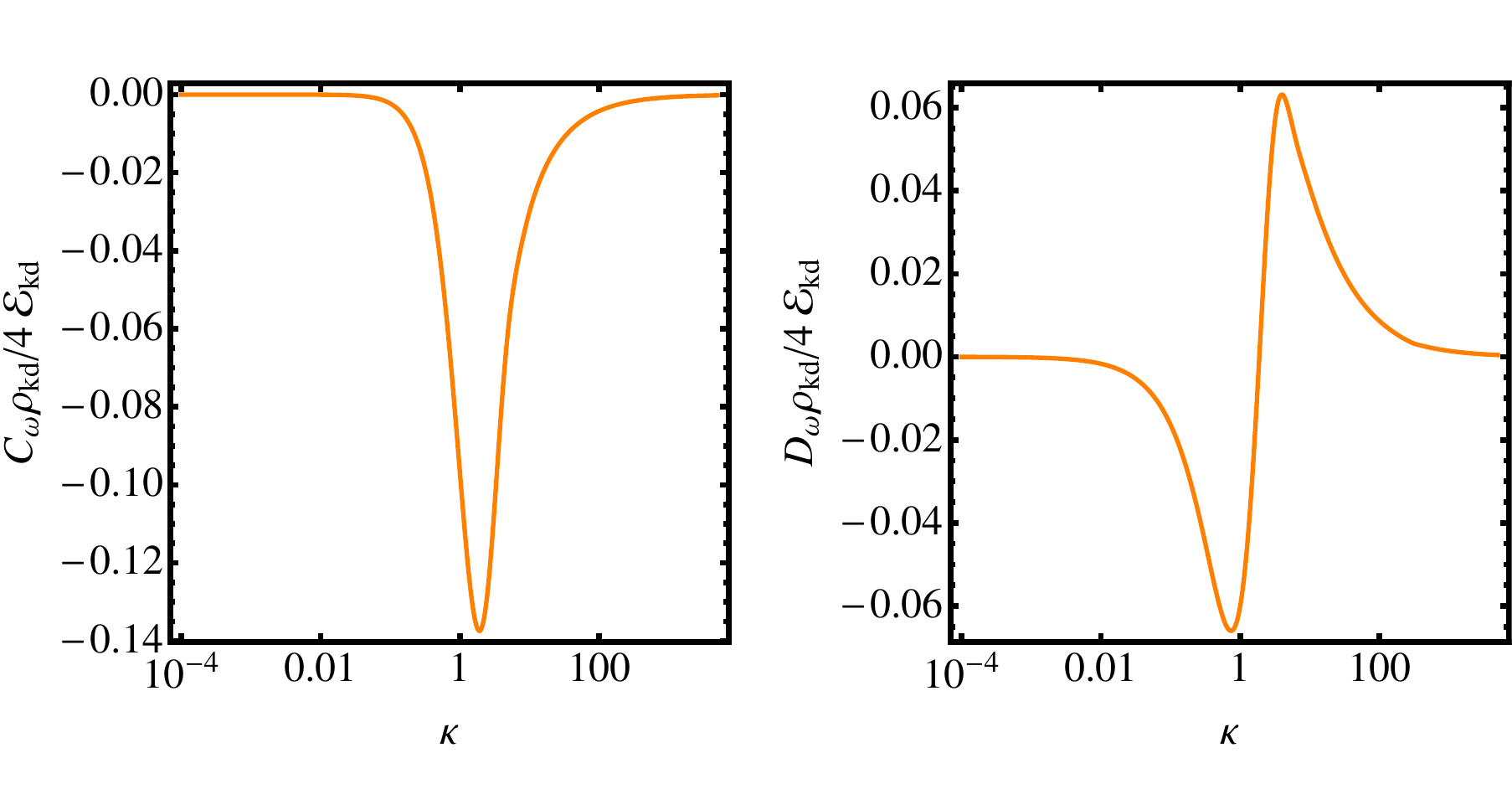}\\
\includegraphics[width=4.6in]{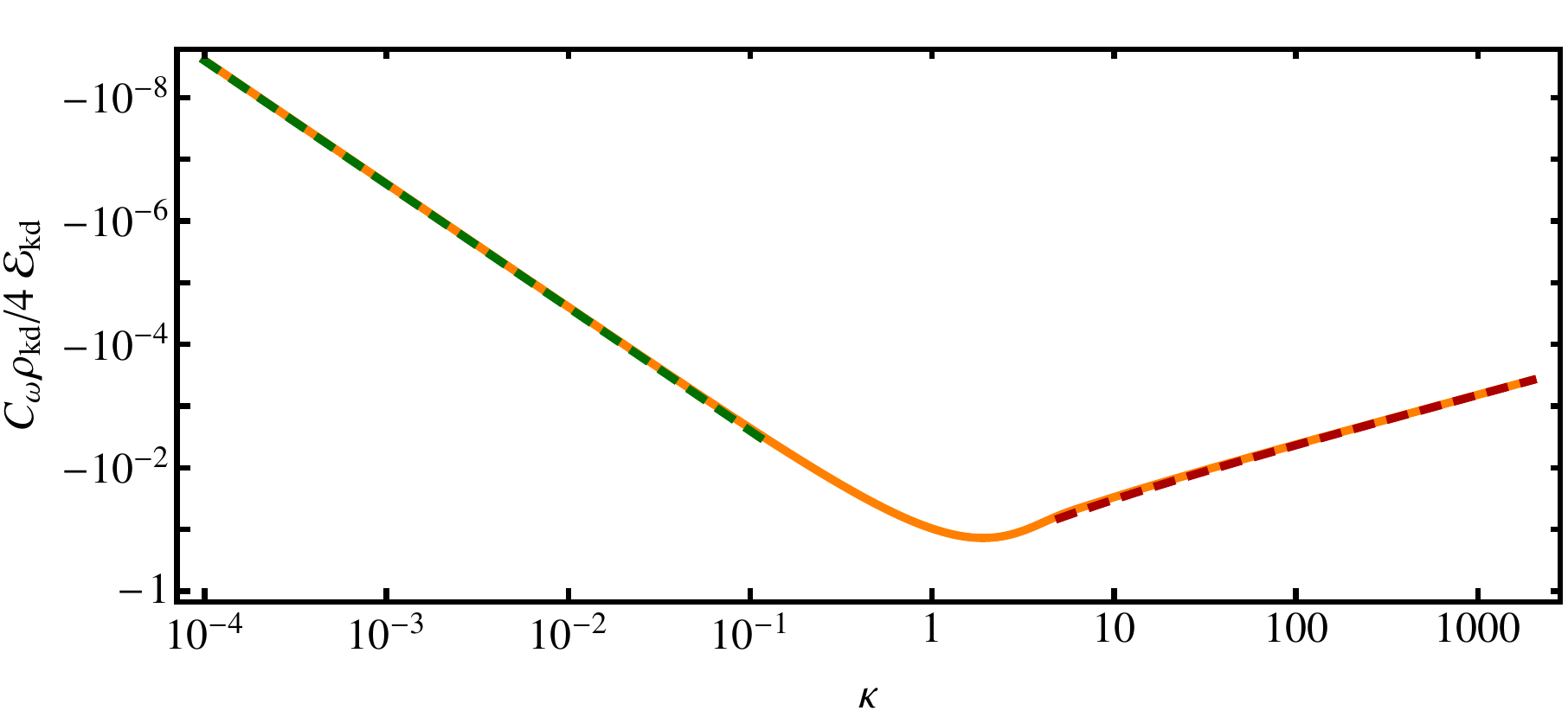}
\caption{Top: Numerical calculation of the damping (left) and phase shift (right) acquired by gravitational waves around the time of kinetic decoupling of the dark matter particles from the standard model. Bottom: Comparison of the numerical computation (orange) with the analytic results described in the text for $\kappa\ll1$ (dashed green) and for $\kappa\gg1$ (dashed red).\label{fig:Ck}}
\end{center}
\end{figure}
Continuing with $\hat\omega(y)= y^{-(4+\beta)}$ for concreteness, we see that the amount of damping experienced around the time when $q/a\approx \omega_{\rm r}$ scales like $\kappa^{-(2+\beta)/(3+\beta)}$,
For $\beta=2$ we, for example, find that gravitational waves with $\kappa\gg1$ are damped by an amount
\begin{equation}
C_\omega\approx-\frac{2\pi\mathcal{E}_{\rm kd}}{5^{5/4}\varphi^{1/2}\rho_{\rm kd}\kappa^{4/5}}\left[1+\frac{\varphi}{\kappa^{4/5}}\right]\qquad{\rm with}\qquad\varphi=\frac{1+\sqrt{5}}{2}\,.
\end{equation}

This result is compared with a numerical calculation in Figure~\ref{fig:Ck}. We see that the power spectrum of primordial gravitational waves carries information both about when kinetic decoupling occurs and about the type of interactions of the dark matter with the standard model. 

Our discussion did not crucially rely on the assumption that the collisions are between dark matter particles and standard model particles and readily extends to models of interacting dark matter. In the presence of dark matter self-interactions, $\hat{\omega}$ in the exponentials of equations~(\ref{eq:Ckd2}) and~(\ref{eq:Dkd2}) should be replaced by the total rate at which collisions transfer energy between dark matter particles, either by collisions with the standard model particles or by self-interactions. Elastic self-interactions do not affect the temperature evolution, and the rate in equation~(\ref{eq:taudm}) that controls the dark matter temperature evolution remains the rate associated with elastic interactions with the standard model unless there are number changing interactions in the dark sector, such as $3\to 2$ processes, or the dark sector contains several degrees of freedom. 

Self-interactions lead to additional collisions which will isotropize the distribution function of dark matter particles more rapidly. This reduces the anisotropic stress and the effect of dark matter on gravitational waves. Besides this general expectation, any discussion of dark matter self-interactions is highly model-dependent, and we will not attempt to classify all possible models. Instead, we content ourselves with a simple concrete example to illustrate that self-interactions may also leave imprints on the gravitational wave spectrum, and imagine a scenario in which the dark matter undergoes elastic self-interactions. The thermally averaged cross-section for elastic scattering of non-relativistic particles is constant, leading to a contribution to the relaxation rate that redshifts like the density of dark matter particles, $y^{-3}$. As we saw earlier, the contribution to the relaxation rate from interactions with standard model particles redshifts faster by at least one power of $y$. For example, if interactions between the dark matter and the standard model are controlled by a four-fermion interaction, they lead to a contribution to the relaxation rate that redshifts like $y^{-6}$. Here three powers of the scale factor arise because the density of standard model particles redshifts like $y^{-3}$, two powers arise from the thermally averaged cross section, and the last power of the scale factor arises because it takes $m/T$ collisions to transfer energies of order $T$ in collisions of the non-relativistic dark matter with the relativistic dark matter particles. 
After the dark matter has frozen out, the number density of standard model particles is exponentially larger than the number density of dark matter particles so that the relaxation rate would presumably initially be dominated by scattering of the dark matter particles with standard model particles. However, because the contribution to the relaxation rate from collisions with the standard model particles redshifts more rapidly as the universe expands, the contributions from dark matter self-interactions would dominate below a certain temperature. The power spectrum of primordial gravitational waves would then contain information about the interactions with the standard model particles or the self-interactions depending whether $q/a\approx \omega$ when the interactions with the standard model particles or the self-interactions dominate the relaxation rate. In this example the evolution of the dark matter temperature remains unchanged, and our results such as~(\ref{eq:Cxan}),~(\ref{eq:Dxan}) directly apply. As long as $\hat{\omega}$ is a superposition of power laws, away from the transition region even the scaling of $C_\omega$ with $\kappa$ we derived for a single power law can be used. In models that modify the evolution of the dark matter temperature some additional work is required, but this is in principle straightforward as well. We see that gravitational waves carry a great deal of information about the properties of dark matter. The only problem is that the effects are hopelessly small. 

We are now also in a position to justify the statement we made in our discussion of long modes for which $\kappa\ll1$, namely that the change in amplitude and phase acquired around the time of kinetic decoupling are much smaller than the contributions acquired after kinetic decoupling. To see this we consider the behavior of the amplitude and phase at a time after kinetic decoupling, but early enough so that the modes are still outside the horizon because this is when we started the computation for the long modes. At this time the arguments of the trigonometric functions for long modes are small and equations~(\ref{eq:Ckd2}) and~(\ref{eq:Dkd2}) become
\begin{eqnarray}
C(x)&\approx&\hphantom{-}\frac{4\mathcal{E}_{\rm kd}\kappa^2}{3\rho_{\rm kd}}\int_{x_1}^x dy\, \tau_{\rm dm} (y)\int_{0}^y dz z\exp\left[-\int_z^y dz' z' \hat{\omega}(z')\right]\,,\\
D(x)&\approx&-\frac{4\mathcal{E}_{\rm kd}\kappa^3}{3\rho_{\rm kd}}\int_{x_1}^x dy\, y\tau_{\rm dm} (y)\int_{0}^y dz z\exp\left[-\int_z^y dz' z' \hat{\omega}(z')\right]\,.
\end{eqnarray}
To make contact with our discussion of long modes, we need $C(x)$ and $D(x)$ sufficiently long after decoupling but before horizon entry. We can write them as
\begin{equation}
C(x)=C_\omega+\frac{2\mathcal{E}_{\rm kd}\tau_{\rm kd}\kappa^2}{3\rho_{\rm kd}}x\qquad\text{and}\qquad D(x)=D_\omega(x)-\frac{\mathcal{E}_{\rm kd}\tau_{\rm kd}\kappa^3}{3\rho_{\rm kd}}x^2\,,
\end{equation}
where $C_\omega$ and $D_\omega(x)$ are given by
\begin{eqnarray}
C_\omega&=&\hphantom{-}\frac{4\mathcal{E}_{\rm kd}\kappa^2}{3\rho_{\rm kd}}\int_{x_1}^x dy\, \tau_{\rm dm} (y)\int_{0}^y dz z\exp\left[-\int_z^y dz' z' \hat{\omega}(z')\right]-\frac{2\mathcal{E}_{\rm kd}\tau_{\rm kd}\kappa^2}{3\rho_{\rm kd}}x\,,\\
D_\omega(x)&=&-\frac{4\mathcal{E}_{\rm kd}\kappa^3}{3\rho_{\rm kd}}\int_{x_1}^x dy\, y\tau_{\rm dm} (y)\int_{0}^y dz z\exp\left[-\int_z^y dz' z' \hat{\omega}(z')\right]+\frac{\mathcal{E}_{\rm kd}\tau_{\rm kd}\kappa^3}{3\rho_{\rm kd}}x^2\,.\label{eq:Domegalong}
\end{eqnarray}
Here $\tau_{\rm kd}$ is defined through the behavior of the dark matter temperature at late times, which according to equation~(\ref{eq:taudm}) is 
\begin{equation}
\tau_{\rm dm}(y)\to \frac{\tau_{\rm kd}}{y^2}\quad{\rm for}\quad y\gg1\,.\label{eq:Tdmasymp}\\
\end{equation}

To see that $C_\omega$ is indeed independent of $x$, note that as the argument of the exponential after kinetic decoupling approaches unity, the terms linear in $x$ cancel, and the remainder is finite. As we mentioned in our discussion of long modes, the term in $C(x)$ linear in $x$ ensures that there is no dependence on the time at which we match onto the collisionless description. Unlike for intermediate modes for which the dominant contribution to $C_\omega$ arises when $q/a\sim \omega_r$, the dominant contribution here arises around kinetic decoupling, and we see that $C_\omega$ universally scale like $\kappa^2$.

The additional factor of $y$ in the integral for $D_\omega(x)$, introduces a logarithmic dependence on $x$ that is absent in the collisionless description. As a consequence, unlike $C_\omega$, the phase receives contributions until horizon crossing. Equation~(\ref{eq:Domegalong}) implies that the contribution from the time around kinetic decoupling universally scales like $\kappa^3$. The presence of two powers of $\kappa$ in the denominators of the mode functions in~(\ref{eq:Dkd2}) implies that the contribution from horizon entry scales like $\kappa$ and dominates.

In the model with $\hat{\omega}=y^{-(4+\beta)}$, the solution to equation~(\ref{eq:taudm}) can be found explicitly in terms of incomplete $\Gamma$-functions and by taking the late time limit we see that the constant in equation~(\ref{eq:Tdmasymp}) is given by 
\begin{equation}
\tau_{\rm kd}=(2+\beta)^{-\frac{1}{2+\beta}}\Gamma\left(\frac{1+\beta}{2+\beta}\right)\,.
\end{equation}
Approximating the integrand of the $y$-integral in $C_\omega$ by its asymptotic forms
\begin{equation}
y^{3+\beta}\quad\text{for}\quad y\leq y_c\,,\qquad\text{and}\qquad\frac12\tau_{\rm kd}\left(1-\frac{(2+\beta)^{-\frac{2}{2+\beta}}\Gamma\left(\frac{\beta}{2+\beta}\right)}{y^2}\right)\quad\text{for}\quad y> y_c\,,
\end{equation}
with
\begin{equation}
y_c=\left(\frac{\tau_{\rm kd}}{2}\right)^{\frac{1}{3+\beta}}\,,
\end{equation}
we find 
\begin{eqnarray}
C_\omega&=&-\frac{2\mathcal{E}_{\rm kd}\tau_{\rm kd}\kappa^2}{3\rho_{\rm kd}}\left[\frac{3+\beta}{4+\beta}\left(\frac{\tau_{\rm kd}}{2}\right)^{\frac{1}{3+\beta}}+(2+\beta)^{-\frac{2}{3+\beta}}\left(\frac{2}{\tau_{\rm kd}}\right)^{\frac{1}{3+\beta}}\Gamma\left(\frac{\beta}{2+\beta}\right)\right]\,.\\
\end{eqnarray}
The phase $D_\omega(x)$ can be  evaluated in the same way, but as we discussed the contribution from kinetic decoupling is suppressed by two powers of $\kappa$ compared to the dominant contribution arising at horizon crossing and we will not give it here.

The variables used here and in the discussion of the long modes are related according to
\begin{equation}
\epsilon\varkappa=\frac{\mathcal{E}_{\rm kd}\tau_{\rm kd}\kappa}{\rho_{\rm kd}}\,.
\end{equation}
For $u_\star=\kappa x_\star \ll1$ the constants $A$ and $B$ in equation~(\ref{eq:hAB}) can then be written as
\begin{equation}
A\approx 1+\epsilon\varkappa \frac{2u_\star}{3}+C_\omega\qquad\text{and}\qquad B\approx-\epsilon\varkappa \frac{u_\star^2}{3}\,.
\end{equation}

For modes that obey $(q/a)v/H\ll1$ around the time of kinetic decoupling so that \mbox{$\kappa\lesssim 1/v_{\rm kd}$}, equations~(\ref{eq:Ckd2}) and~(\ref{eq:Dkd2}) and our discussion here are valid throughout. For a typical WIMP this corresponds to frequencies of below $\sim 10^{-9}$ Hz today. For modes with shorter wavelengths we must understand whether higher multipoles may become excited. To gain some intuition we will make the simplifying assumption that the relaxation rates for all $n$ and $\ell$ are identical to those for $n=0$ and $\ell=2$. This is equivalent to working in the relaxation time approximation. In this case the derivation from section III goes through essentially unchanged and the anisotropic stress is given by
\begin{eqnarray}
&&\pi_{q}(t)=\frac{\pi}{4ma^5(t)}\int_0^\infty p^5\,dp\; \overline{n}'(p)\int_{-1}^{+1} (1-\mu^2)^2\,d\mu\nonumber\\&&~~~~~~\times\int_{t_1}^{t}dt'\dot{h}_{q}(t')\exp\left[-
\int^t_{t'}dt''\frac{iqp\mu}{a^2(t'')m}\right] \exp\left[-2\int^t_{t'}dt''\omega_{\rm r}(t'')\frac{T(t'')}{T_{\rm dm}(t'')}\right]\;.\label{eq:pi_int}
\end{eqnarray}
As before, the equation of motion at late times when $q/a\gg \omega_{\rm r}, H$ is given by equation~(\ref{eq:heom_hom}), and we only have to follow the evolution of the mode until $q/a\gg \omega_{\rm r}\gg H$ to find the appropriate initial conditions for this equation.

To find the expression for the anisotropic stress valid from horizon entry until $q/a\gg \omega_{\rm r}\gg H$, we can proceed as before and approximate the second line of equation~(\ref{eq:pi_int}) as
\begin{equation}
\int_{0}^y dz h_q^{1\prime}(z)\exp\left[-\frac{i\kappa p\mu}{m a_{\rm kd}}\ln\frac{y}{z}-\int_z^y dz' z' \hat{\omega}(z')\right]\approx\int_{-\infty}^0 du h^{1\prime}_q(y+u)\exp\left[i\frac{\kappa p\mu u}{m a}\right]\exp\left[uy\hat\omega(y)\right]\,.
\end{equation}
The integral on the right hand side only receives significant contributions for $|u|<1/\kappa$ so that the argument of the first exponential is of order the dark matter velocity $v$ around the time when $q/a\approx \omega_{\rm r}$. Furthermore, because of the integration over $\mu$ in equation~(\ref{eq:pi_int}) only even powers in $\mu$ contribute so that the leading correction occurs at second order in the dark matter velocity, implying that the damping of the amplitude and the phase shift for all intermediate modes are well approximated by equations~(\ref{eq:Cxan}) and~(\ref{eq:Dxan}). Furthermore, for all modes that enter after the dark matter particles have become non-relativistic, $q/a\approx\omega_{\rm r}$ occurs after freeze-out so that annihilations can be neglected around this time. 

\paragraph{Short modes}\mbox{}\\
We now turn to modes that enter the horizon when the dark matter is still relativistic. While detailed modeling of the collision terms describing the scattering of relativistic dark matter particles with the standard model is possible, it is significantly more tedious than in the non-relativistic limit, and we continue with the simplifying assumption that relaxation rates for all $n$ and $\ell$ are equivalent to those for $n=0$ and $\ell=2$. In this case the anisotropic stress is 
\begin{eqnarray}
&&\hskip -1cm\pi_q(t)=\frac{\pi}{4a^5(t)}\int_0^\infty p^5\,dp\; \frac{\overline{n}'(p)}{\sqrt{m^2+p^2/a^2(t)}}\int_{-1}^{+1} (1-\mu^2)^2\,d\mu\nonumber\\
&&\times\int_{t_1}^{t}dt'\dot{h}_q(t')\exp\left[-
\int^t_{t'}dt''\frac{iqp\mu}{a^2(t'')\sqrt{m^2+
p^2/a^2(t'')}}\right]\exp\left[-2\int^t_{t'}dt''\omega(t'')\right]\;,\label{eq:pi_rel}
\end{eqnarray}
with $\omega(t)$ now the collision rate including both elastic an inelastic processes. 

Short modes naturally subdivide into two classes, one for which the dark matter is still relativistic and one for which it is non-relativistic when $q/a\approx\omega$. For a typical WIMP, the boundary between these classes corresponds to modes with a frequency of $10^{4}$ Hz today, so that for all planned interferometer experiments it is sufficient to focus on modes for which the dark matter is already non-relativistic when $q/a\approx\omega$. As we will see, the dominant contributions for these modes arise during two periods, the first around the time when the dark matter becomes non-relativistic, and the second when $q/a\approx\omega$. Scattering is very rapid during both periods and we expect~(\ref{eq:pi_rel}) to provide a very good approximation.

From the discussion of intermediate modes, we know that the equation of motion for gravitational waves when $q/a\gg \omega, qv/a$ is given by equation~({\ref{eq:heom_hom}}). What remains is to find the initial conditions for this equation or equivalently the amplitude and phase shift. As before we will make use of the fact that the dark matter distribution approaches its equilibrium value on time scales short compared to the expansion of the universe and the integral over $t'$ receives its dominant contribution near the upper limit. Using the same notation as for the intermediate modes, we can approximate 
\begin{eqnarray}
&&\hskip -1.5cm\int_{0}^y dz h_q^{1\prime}(z)\exp\left[-i\int_z^y dz'\frac{i\kappa p\mu}{ z' a_{\rm kd}\sqrt{m^2+p^2/z^{\prime 2}/a_{\rm kd}^2}}-\int_z^y dz' z' \hat{\omega}(z')\right]\\
&&\hskip 1.5cm\approx\int_{-\infty}^0 du h^{1\prime}_q(y+u)\exp\left[i\frac{\kappa p\mu u}{a\sqrt{m^2+p^2/a^2}}\right]\exp\left[uy\hat\omega(y)\right]\,.
\end{eqnarray}
The integral over $u$ only receives significant contributions for $|u|$ of order $1/y\hat\omega(y)$, which is of order $1/\kappa$ when $q/a\approx \omega$. This implies that the argument of the argument of the exponential is of order the dark matter velocity at this time and hence small for the modes of interest. The integration over $\mu$ implies that the leading contribution arises at second order in the velocities and we will ignore these corrections. At earlier times $y\hat\omega(y)\gg\kappa$ so that the argument is further suppressed then, and we can approximate the anisotropic stress by
\begin{eqnarray}
&&\hskip -1cm\pi_q(t)=\frac{4\pi}{15a^5(t)}\int_0^\infty p^5\,dp\; \frac{\overline{n}'(p,t)}{\sqrt{m^2+p^2/a^2(t)}}\int_{-\infty}^0 du h^{1\prime}_q(y(t)+u)\exp\left[uy(t)\hat\omega(y(t))\right]\;.
\end{eqnarray}
As long as $y\hat\omega\gg \kappa$, which is the case for the short modes of interest until the dark matter has become non-relativistic, we can neglect $u$ in $h_q^{1\prime}$ and the equation of motion for gravitational waves becomes
\begin{eqnarray}
 h_q''(x)+\left(\frac{2}{x}+\gamma(x)\right)h_q'(x)+\kappa^2 h_q(x)=0\,,
\end{eqnarray}
with
\begin{eqnarray}
\gamma(x)=\frac{2}{5 \rho(x) x^3\hat\omega(x)}\int \frac{d^3 p}{(2\pi)^3}\frac{p^2(4E^2+m^2)}{(xa_{\rm kd})^5E^3}\overline{n}(p,t)\;,
\end{eqnarray}
where $\rho(x)$ is the total energy density and $E=\sqrt{m^2+p^2/a^2}$. Either treating the additional damping term as a perturbation and using the Green's function~(\ref{eq:Greens_rad}) or using the WKB approximation, we find that the damping of the amplitude is independent of wave number and is given by
\begin{equation}
C(x)=-\frac{4}{5}\int_0^x dy\frac{f_{\rm dm}(y)}{ y^3\hat\omega(y)}\qquad{\rm with}\qquad f_{\rm dm}(y)=\frac{1}{4\rho(y)}\int d^3 p\frac{p^2(4E^2+m^2)}{(ya_{\rm kd})^5E^3}\overline{n}(p,t)\,.
\end{equation}
At early times when the dark matter is relativistic, $f_{\rm dm}$ is time-independent and corresponds to the fraction of the energy density stored in dark matter. As the temperature drops below the mass of the dark matter particles, $f_{\rm dm}(y)$ decreases rapidly and cuts off the integral. In general, $\overline{n}(p,t)$ follows from the freeze-out calculation based on equation~(\ref{eq:nfreeze}). For scattering rates that do not drop too rapidly, we can approximate $\overline{n}(p,t)$ by its equilibrium abundance and write
\begin{equation}
f_{\rm dm}(y)=\frac{g_d}{g_\star(y)}\frac{30}{\pi^2}\int_0^\infty \frac{dz}{2\pi^2} \frac{z^4(5s^2/4+z^2)}{(s^2+z^2)^{3/2}}\frac{1}{e^{\sqrt{s^2+z^2}}\pm 1}\qquad{\rm with}\qquad s=\frac{m}{T}=\frac{m}{T_{\rm kd}}y\,,
\end{equation}
with $g_d$ counting the number of degrees of freedom in the dark matter, and $g_\star(y)$ the usual effective number of degrees of freedom.

If the interactions between the dark matter particles and the standard model are controlled by a single operator, we expect $\hat\omega(y)=(m/T_{\rm kd})y^{-(3+\beta)}$. In this case the amount of damping experienced around the time when the dark matter becomes non-relativistic is given by
\begin{equation}
C_{\rm nr}=-\frac{4}{5}\frac{g_d}{g_{\star,m}}\left(\frac{T_{\rm kd}}{m}\right)^{2+\beta}\mathcal{F}(\beta)\,,
\end{equation}
where $g_{\star,m}$ is the effective number of relativistic degrees of freedom around the time when the dark matter particles become non-relativistic, and $\mathcal{F}(\beta)$ is a function that only depends on $\beta$ and can readily be evaluated numerically. In our example of $\beta=2$, it takes the value $\mathcal{F}(2)\approx 12$. For larger values of $\beta$, $\overline{n}(p,t)$ should be obtained using equation~(\ref{eq:nfreeze}). Since $T_{\rm kd}/m$ is the square of the dark matter velocity at kinetic decoupling, we see that the effect is rather small.

We now know the mode functions for short modes at a time when $q/a$ is still small compared to $\omega$ but the dark matter has already become non-relativistic. We can proceed just like for the intermediate modes to evolve the modes until $q/a\gg\omega$ and equation~(\ref{eq:heom_hom}) describes their evolution. The only difference is that for intermediate modes the lower limit of the integral in equation~(\ref{eq:intappr}) was zero whereas it is now non-zero. However, the integral is dominated by the contribution near the upper limit so that this difference is negligible and the damping and phase shift acquired around the time when $q/a\approx \omega_{\rm r}$ are still given by equations~(\ref{eq:Cxan}) and~(\ref{eq:Dxan}).

As long as the two events are separated, the total amount of damping is simply given by $C_{\rm nr}+C_{\rm \omega}$. For high frequencies the first term dominates, for low frequencies it is the second. Up to order one factors, the transition between the regime occurs at
\begin{equation}
\kappa\sim\left(\frac{g_{\star,{\rm eq}}}{g_d}\right)^{\frac{3+\beta}{2+\beta}}\left(\frac{g_{\star,{\rm m}}}{g_{\star, {\rm kd}}}\right)^\frac{3+\beta}{2+\beta}\left(\frac{T_{\rm eq}}{m}\right)^\frac{3+\beta}{2+\beta}\left(\frac{m}{T_{\rm kd}}\right)^{3+\beta}\,,
\end{equation}
with frequency independent damping above this wave number and an amount of damping that scales like $k^{-(2+\beta)/(3+\beta)}$ for smaller wave numbers.

\section{Conclusions}
We have analyzed the effects of cold dark matter on the propagation of gravitational waves of astrophysical and primordial origin. Our analysis does not suggest any way of detecting the effect of cold dark matter on the propagation of gravitational waves from astrophysical gravitational waves in the near future. 

Primordial gravitational waves in principle contain a wealth of information about dark matter and its interactions such as coupling strengths and the nature of the interactions. However, in practice the effects of cold dark matter on primordial gravitational waves also appear too small to be detectable. For the longest modes that enter after matter radiation equality, the anisotropic stress is small because the cold dark matter is highly non-relativistic by the time of horizon entry. The effects are largest for intermediate modes that enter the horizon around the time of kinetic decoupling, but even then the effects are highly suppressed because the cold dark matter is non-relativistic at this time and because the contribution to the energy density from dark matter is small compared to that in radiation at the time of kinetic decoupling.  For shorter modes, the effects are suppressed because collisions rapidly drive the system toward local equilibrium. 

Unlike cold dark matter, particles that decouple when they are relativistic have a significant effect on primordial gravitational waves. Modes that enter after kinetic decoupling are damped~\cite{Weinberg:2003ur}. The spectrum of primordial gravitational waves on scales that enter the horizon around the time of kinetic decoupling contains information about the interactions. However, for neutrinos, the only particles known to decouple relativistically, kinetic decoupling is imprinted on modes with frequencies that are too high to be accessible in the CMB and too low for pulsar timing arrays.

\newpage
\begin{center}
Acknowledgments
\end{center}
We are grateful for helpful conversations with Richard Matzner and Paul Shapiro. R.F. was supported in part by the Alfred P. Sloan Foundation, the Department of Energy under Grant No. DE-SC0009919, and a grant from the Simons Foundation/SFARI 560536. S.W. was supported by the National Science Foundation under Grant Number PHY-1620610 and by The Robert A. Welch Foundation, Grant No. F-0014.
\vspace{10pt}
\section*{APPENDIX A: Boltzmann Hierarchy}
In this appendix we provide the derivation of the Boltzmann hierarchy~(\ref{eq:Dnann}) from the linearized Boltzmann equation~(\ref{eq:linearBEs}). As we explained in section VII, the form of the mode expansion for the gravitational field given in equation~(\ref{eq:h_mode}) implies that equation~(\ref{eq:linearBEs}) only depends on the direction of the momentum of the dark matter particles through $\mu=\hat{p}\cdot\hat{q}$ and $e_{ij}(\hat{q},\lambda)\hat{p}^i\hat{p}^j$. For isotropic initial conditions, the perturbation to the phase space density of the dark matter particles introduced by the gravitational wave must then be of the form~(\ref{eq:dnchiansatz}).
For this Ansatz equation~(\ref{eq:linearBEs}) becomes a differential equation for $\tilde\Delta(q,p,\mu,t)$ 
\begin{eqnarray}\label{eq:linearBE4}
&&\hskip -1cm\dot{\tilde\Delta}(q,p,\mu,t)+\frac{i p q \mu}{a^2 m}\tilde\Delta(q,p,\mu,t)-\frac12\dot{h}_q(t)p\frac{\partial}{\partial p}\overline{n}(p,t)=\nonumber\\
&&\hskip 1cm-2\omega_{\rm a}(t)\tilde\Delta(q,p,\mu,t)+\omega_{\rm r}(t)\left[3 \tilde\Delta(q,p,\mu,t)+p\frac{\partial}{\partial p}\tilde\Delta(q,p,\mu,t)\right.\nonumber\\
&&\hskip 5.5cm\left.+a^2m T\left(\frac{\partial^2}{\partial p^2}+\frac2p\frac{\partial}{\partial p}-\frac{1}{p^2}\mathcal{D}^2\right)\tilde\Delta(q,p,\mu,t)\right]\,,
\end{eqnarray}
with the operator $\mathcal{D}^2$ given by
\begin{equation}
\mathcal{D}^2=-(1-\mu^2)\frac{\partial^2}{\partial \mu^2}+6\mu\frac{\partial}{\partial \mu}+6\,.
\end{equation}
We will eventually expand in terms of eigenfunctions of $\mathcal{D}^2$ and the differential operator appearing on the right hand side. Since it involves $T$ instead of $T_{\rm dm}$, one would have to keep a large number of the eigenfunctions when $T_{\rm dm}\ll T$. In an attempt to ameliorate this, we will work with the fractional perturbation $\Delta(q,p,\mu,t)$ defined by
\begin{equation}
\tilde\Delta(q,p,\mu,t)=\Delta(q,p,\mu,t)p\frac{\partial}{\partial p}\overline{n}(p,t)\,.
\end{equation}
For simplicity, let us drop the first term on the right hand side because $\omega_{\rm a}\ll \omega_{\rm r}$ when the dark matter particles are non-relativistic. We will restore it later. In this case the equation becomes
\begin{eqnarray}\label{eq:linearD}
&&\hskip -1.2cm\dot{\Delta}(q,p,\mu,t)+\frac{i p q \mu}{a^2 m}\Delta(q,p,\mu,t)-\frac12\dot{h}_q(t)=\nonumber\\
&&\hskip -0.4cm\omega_{\rm r}(t)\left[-\frac{2T}{T_{\rm dm}}\Delta(q,p,\mu,t)-\left(\frac{2T}{T_{\rm dm}}-1\right)p\frac{\partial}{\partial p}\Delta(q,p,\mu,t)\nonumber\right.\\
&&\hskip 1.0cm\left.+a^2m T\left(\frac{\partial^2}{\partial p^2}+\frac6p\frac{\partial}{\partial p}+\frac{6}{p^2}-\frac{1}{p^2}\mathcal{D}^2\right)\Delta(q,p,\mu,t)\right].
\end{eqnarray}
Our goal will be to turn this partial differential equation into a hierarchy of coupled ordinary differential equations by constructing the eigenfunctions of the differential operator on the right hand side and rely on the orthogonality of eigenfunctions with different eigenvalues. The eigenfunctions of the operator $\mathcal{D}^2$ with appropriate boundary conditions are 
\begin{equation}
\mathcal{P}_\ell(\mu)=\frac{P_\ell^2(\mu)}{1-\mu^2}\,,
\end{equation}
where $P_\ell^m(\mu)$ are associated Legendre polynomials. These functions are eigenfunctions of $\mathcal{D}^2$ with eigenvalue $\ell(\ell+1)$
\begin{equation}
\mathcal{D}^2\mathcal{P}_\ell(\mu)=\ell(\ell+1)\mathcal{P}_\ell(\mu)\,,
\end{equation}
and obey the orthogonality relation
\begin{equation}\label{eq:orth1}
\int_{-1}^1 d\mu (1-\mu^2)^2\mathcal{P}_\ell(\mu)\mathcal{P}_{\ell'}(\mu)=\frac{2(\ell+2)!}{(2\ell+1)(\ell-2)!}\delta_{\ell\ell'}\,.
\end{equation}
For $\ell=2$ we simply have
\begin{eqnarray}
\mathcal{P}_2(\mu)&=&3\,,
\end{eqnarray}
so that the orthogonality relation also implies
\begin{equation}\label{eq:orth2}
\int_{-1}^1 d\mu (1-\mu^2)^2\mathcal{P}_\ell(\mu)=\frac{16}{5}\delta_{\ell2}\,.
\end{equation}
Furthermore, they obey the recurrence relation
\begin{equation}\label{eq:rec1}
\mu\mathcal{P}_\ell(\mu)=\frac{\ell+1}{2\ell+1}\mathcal{P}_{\ell-1}(\mu)+\frac{\ell-1}{2\ell+1}\mathcal{P}_{\ell+1}(\mu)\,.
\end{equation}
Expanding
\begin{equation}
\Delta(q,p,\mu,t)=\sum_\ell(-i)^\ell(2\ell+1)\Delta_\ell(q,p,t)\mathcal{P}_\ell(\mu)\,,
\end{equation}
and using the recurrence relation~(\ref{eq:rec1}), the orthogonality relations~(\ref{eq:orth1}) and~(\ref{eq:orth2}), equation~(\ref{eq:linearD}) becomes
\begin{eqnarray}\label{eq:linearE}
&&\hskip -1.2cm\dot{\Delta}_\ell(q,p,t)+\frac{p q}{(2\ell+1)a^2 m}\Big[(\ell+2)\Delta_{\ell+1}(q,p,t)-(\ell-2)\Delta_{\ell-1}(q,p,t)\Big]+\frac{1}{30}\dot{h}_q(t)\delta_{\ell,2}=\nonumber\\
&&\hskip 2.6cm\omega_{\rm r}(t)\left[-\frac{2T}{T_{\rm dm}}\Delta_\ell(q,p,t)-\left(\frac{2T}{T_{\rm dm}}-1\right)p\frac{\partial}{\partial p}\Delta_\ell(q,p,t)\right.\nonumber\\
&&\hskip 4cm\left.+a^2m T\left(\frac{\partial^2}{\partial p^2}+\frac6p\frac{\partial}{\partial p}-\frac{\ell(\ell+1)-6}{p^2}\right)\Delta_\ell(q,p,t)\right].
\end{eqnarray}
It would seem natural to work with the eigenfunctions of the operator on the right hand side. However, it turns out to be convenient to instead work with the eigenfunctions  
\begin{eqnarray}
&&\hskip -2cm\left[-\frac{T}{T_{\rm dm}} p\frac{\partial}{\partial p}+a^2m T\left(\frac{\partial^2}{\partial p^2}+\frac6p\frac{\partial}{\partial p}-\frac{\ell(\ell+1)-6}{p^2}\right)\right]\mathcal{L}_{n\ell}(z)=\nonumber\\
&&\hskip 6cm -(2n+\ell-2)\frac{T}{T_{\rm dm}}\mathcal{L}_{n\ell}(z)\,,
\end{eqnarray}
with $n=0\dots\infty$ and $\ell=2\dots\infty$, which are given in terms of generalized Laguerre polynomials $L_n^k$ by 
\begin{equation}
\mathcal{L}_{n\,\ell}(z)=z^{\ell/2-1}L_n^{\ell+1/2}(z)\qquad{\rm with}\qquad z=\frac{p^2}{2a^2m T_{\rm dm}}\,.
\end{equation}
As we will see, the advantage of this basis is that $z$ is also the argument of the exponential in $\overline{n}(p,t)$. These functions obey the orthogonality relation
\begin{equation}\label{eq:orthL}
\int_0^\infty dz\,z^{5/2} e^{-z}\mathcal{L}_{n\,\ell}(z)\mathcal{L}_{n'\,\ell}(z)=\frac{\Gamma(n+\ell+3/2)}{n!}\delta_{nn'}\,,
\end{equation}
which contains the special case
\begin{equation}\label{eq:orthL0}
\int_0^\infty dz\,z^{5/2} e^{-z}\mathcal{L}_{n\,2}(z)=\frac{15\sqrt{\pi}}{8}\delta_{n0}\,.
\end{equation}
To make use of the orthogonality relation~(\ref{eq:orthL}) when deriving the hierarchy, we will have to use the relations
\begin{eqnarray}
\mathcal{L}_{0\,\ell-1}(z)&=& z^{-1/2} \mathcal{L}_{0\,\ell}(z)\,,\\
\frac{d}{dz} \mathcal{L}_{0\,\ell}(z) &=&\frac{\ell-2}{2z}\mathcal{L}_{0\,\ell}(z)\,,\\
\mathcal{L}_{n\,\ell+1}(z)&=&\left(n+\ell+\frac{3}{2}\right) z^{-1/2} \mathcal{L}_{n\,\ell}(z)-(n+1) z^{-1/2}\mathcal{L}_{n+1\,\ell}(z)\,,\\
\mathcal{L}_{n\,\ell-1}(z)&=& z^{-1/2} \mathcal{L}_{n\,\ell}(z)- z^{-1/2}\mathcal{L}_{n-1\,\ell}(z)\qquad\hskip 1.85cm{\rm for}\qquad n\geq 1\,,\\
\frac{d}{dz} \mathcal{L}_{n\,\ell}(z) &=&\frac{2n+\ell-2}{2z}\mathcal{L}_{n\,\ell}(z)-\frac{n+\ell+\frac12}{z}\mathcal{L}_{n-1\,\ell}(z)\qquad{\rm for}\qquad n\geq 1\,,
\end{eqnarray}
which directly follow from the relations for associated Laguerre polynomials
\begin{eqnarray}
L_0^{\ell+1/2}(z)&=&L_0^{\ell+3/2}(z)\,,\\
\frac{d}{dz}L_0^{\ell+1/2}(z)&=&0\,,\\
z L_n^{\ell+3/2}(z)&=&\left(n+\ell+\frac{3}{2}\right)L_n^{\ell+1/2}(z)-(n+1) L_{n+1}^{\ell+1/2}(z)\,,\\
L_n^{\ell+1/2}(z)&=&L_n^{\ell+3/2}(z)- L_{n-1}^{\ell+3/2}(z)\qquad{\rm for}\qquad n\geq 1\,,\\
\frac{d}{dz}L_n^{\ell+1/2}(z)&=&-L_{n-1}^{\ell+3/2}(z)\qquad\hskip 1.85cm{\rm for }\qquad n\geq1\,.
\end{eqnarray}
Expanding $\Delta_\ell(q,p,t)$ in terms of these eigenfunctions
\begin{equation}
\Delta_\ell(q,p,t)=\sum_n \Delta_{n\,\ell}(q,t) \mathcal{L}_{n\,\ell}(z)\,,
\end{equation}
substituting the expansion into equation~(\ref{eq:linearE}), using the orthogonality relation and identities in the appendix, as well as equation~(\ref{eq:Tdmev}) in the form
\begin{equation}
\frac{\dot{z}}{z}=-2\omega_{\rm r}(t)\left(\frac{T}{T_{\rm dm}}-1\right)\,,
\end{equation}
we arrive at the following hierarchy of equations
\begin{eqnarray}
&&\hskip -1.72cm\dot{\Delta}_{n\,\ell}(q,t)+\frac{q}{(2\ell+1)a}\left(\frac{2T_{\rm dm}} {m}\right)^{1/2}\Bigg[(\ell+2)\left(n+\ell+\frac32\right)\Delta_{n\,\ell+1}(q,t)\nonumber\\*
&&-n(\ell+2)\Delta_{n-1\,\ell+1}(q,t)+(\ell-2)\Delta_{n+1\,\ell-1}(q,t)-(\ell-2)\Delta_{n\,\ell-1}(q,t)\Bigg]=\nonumber\\*
&&\hskip 5cm-\frac{1}{30}\dot{h}_q(t)\delta_{\ell2}\delta_{n0}-(2n+\ell)\omega_{\rm r}(t)\frac{T}{T_{\rm dm}}\Delta_{n\,\ell}(q,t)\,.
\end{eqnarray}
The derivation in the presence of annihilations proceeds in the same way, and keeping them one arrives at
\begin{eqnarray}
&&\hskip -1.72cm\dot{\Delta}_{n\,\ell}(q,t)+\frac{q}{(2\ell+1)a}\left(\frac{2T_{\rm dm}} {m}\right)^{1/2}\Bigg[(\ell+2)\left(n+\ell+\frac32\right)\Delta_{n\,\ell+1}(q,t)\nonumber\\*
&&-n(\ell+2)\Delta_{n-1\,\ell+1}(q,t)+(\ell-2)\Delta_{n+1\,\ell-1}(q,t)-(\ell-2)\Delta_{n\,\ell-1}(q,t)\Bigg]=\nonumber\\*
&&\hskip 1cm-\frac{1}{30}\dot{h}_q(t)\delta_{\ell2}\delta_{n0}-(2n+\ell)\omega_{\rm r}(t)\frac{T}{T_{\rm dm}}\Delta_{n\,\ell}(q,t)-2\omega_{\rm a}(t)\frac{n_{{\rm eq}}^2}{n^2}\Delta_{n\,\ell}(q,t)\,.
\end{eqnarray}

\vspace{10pt}

\begin{center}
{\bf ---------}
\end{center}

\end{document}